\documentclass[a4paper]{amsart}

\usepackage{a4wide}
\usepackage[utf8]{inputenc}
\usepackage[T1]{fontenc}
\usepackage{amsmath}
\usepackage{amssymb,empheq,mathtools,dsfont}
\usepackage{physics}
\usepackage{stmaryrd}
\usepackage[mathscr]{eucal}
\usepackage{color}
\usepackage{nicefrac}
\usepackage{thm-restate}
\usepackage{calrsfs}
\usepackage[breaklinks,colorlinks=true, citecolor=blue]{hyperref}
\usepackage{MnSymbol}
\usepackage{enumitem}
\usepackage{graphicx}

\DeclareMathAlphabet{\pazocal}{OMS}{zplm}{m}{n}

\usepackage{caption}
\usepackage{subcaption}

\makeatletter
\def\namedlabel#1#2{\begingroup
	#2\def\@currentlabel{#2}\phantomsection\label{#1}\endgroup
}
\makeatother

\usepackage{scalerel,stackengine}
\stackMath
\newcommand\reallywidehat[1]{\savestack{\tmpbox}{\stretchto{\scaleto{\scalerel*[\widthof{\ensuremath{#1}}]{\kern.1pt\mathchar"0362\kern.1pt}{\rule{0ex}{\textheight}}}{\textheight}}{2.4ex}}\stackon[-6.9pt]{#1}{\tmpbox}}

\newcommand{\C}{\mathbb{C}}
\newcommand{\F}{\mathbb{F}}

\newcommand{\LP}{\textup{LP}}

\renewcommand{\vec}[1]{\mathbf{#1}}

\newcommand{\cv}{\vec{c}}

\newcommand{\xv}{\vec{x}}
\newcommand{\yv}{\vec{y}}

\newcommand{\un}{\vec{1}}
\newcommand{\one}{\un}

\newcommand{\mat}[1]{\mathbf{#1}}

\newcommand{\Dm}{\mat{D}}
\newcommand{\Em}{\mat{E}}

\newcommand{\dmin}{d_{\textup{min}}}

\makeatletter
\newcommand*{\transp}{{\mathpalette\@transpose{}}}
\newcommand*{\@transpose}[2]{\raisebox{\depth}{$\m@th#1\intercal$}}
\makeatother

\newcommand*{\eqdef}{\stackrel{\text{def}}{=}}

\newtheorem{theorem}{Theorem}

\newtheorem{remark}{Remark}
\newtheorem{definition}{Definition}
\newtheorem{proposition}{Proposition}
\newtheorem{lemma}{Lemma}

\newcommand{\X}{\mathsf{X}}

\newcommand{\ie}{{\it i.e.,}}

\newcommand{\hf}{\widehat{f}}
\newcommand{\hg}{\widehat{g}}
\newcommand{\tf}{\widetilde{f}}
\newcommand{\tg}{\widetilde{g}}

\newcommand{\COMMENT}[1]{}
\newcommand{\triple}[3]{\bra{#1}#2\ket{#3}}
\renewcommand{\C}{\mathcal{C}}
\newcommand{\D}{\vec{D}}
\newcommand{\E}{\vec{E}}

\usepackage[foot]{amsaddr}

\title{New Solutions to Delsarte's Dual Linear Programs}

\author{Andr\'{e} Chailloux$^*$}
\address{$*$ Inria de Paris, {\'Equipe COSMIQ}}
\email{andre.chailloux@inria.fr}
\author{Thomas Debris--Alazard$^\dagger$}
\address{$\dagger \hspace{0.04cm}$ Inria Saclay, {\'Equipe GRACE}}
\email{thomas.debris@inria.fr}

\begin{document}
		\begin{abstract}
	Understanding the maximum size of a code with a given minimum distance is a major question in computer science and discrete mathematics. The most fruitful approach for finding asymptotic bounds on such codes is by using Delsarte's theory of association schemes. With this approach, Delsarte constructs a linear program such that its maximum value is an upper bound on the maximum size of a code with a given minimum distance. Bounding this value can be done by finding solutions to the corresponding dual linear program. Delsarte's theory is very general and goes way beyond binary codes. 
	
	In this work, we provide universal bounds in the framework of association schemes that generalize the Elias-Bassalygo bound, which can be applied to any association scheme constructed from a distance function. These bounds are obtained by constructing new solutions to Delsarte's dual linear program. We instantiate these results and we recover known bounds for $q$-ary codes and for constant-weight binary codes. Our other contribution is to recover, for essentially any $Q$-polynomial scheme, MRRW-type solutions to Delsarte's dual linear program which are inspired by the Laplacian approach of Friedman and Tillich
	instead of using the Christoffel-Darboux formulas. We show in particular how the second linear programming bound can be interpreted in this framework.
\end{abstract}

 	\maketitle
	\section{Introduction}

Let $\tau_{\textup{H}}$ denote the Hamming distance. 
For a subset of the boolean cube, \ie{} a binary code $\C \subseteq \F_2^n$, its minimum distance is~$d^{\textup{H}}_{\textup{min}}(\C) \eqdef \min\{\tau_{\textup{H}}\left( \cv - \cv' \right):\; \cv,\cv' \in \C \textrm{ with } \cv \neq \cv'\}$. We define $A(n,d)$ as being the maximum size of a binary code with some fixed minimum distance, \ie{}
$$ 
A(n,d) \eqdef \max\left\{ |\C| :\; \C \subseteq \F_2^n, \ d^{\textup{H}}_{\textup{min}}(\C) = d\right\}.
$$
We are interested in the maximum asymptotic rate of binary codes with a certain (relative) minimum distance~$\delta \in [0,1]$, \ie{}
$$ 
R(\delta) \eqdef \mathop{\overline{\lim}}\limits_{n \rightarrow \infty}\frac{1}{n} \log_{2}A(n,\lfloor \delta n \rfloor).
$$
There are many reasons why one can be interested in this quantity. Constructing binary codes with large minimum distance is studied since the seminal work of Shannon~\cite{Sha48}. Understanding~$R(\delta)$ has important consequences for telecommunications, in particular to provide lower bounds on the probability of undetected error and for finding optimal codes for error detection over the binary-symmetric channel~\cite{SGB67}. Moreover, it is a fundamental question in discrete mathematics and it is strongly related to sphere packings in $\mathbb{R}^n$~\cite{KL78,CE03}. Finally, the fact that the best bounds on~$R(\delta)$ were found in the late $1970$'s -  which are quite far from what we expect - makes it a particularly interesting question for mathematicians and computer scientists. 

Another important question is to provide bounds on the size of constant-weight codes with a certain minimum distance. These codes are subsets of $\mathcal{S}_{a}^{n,2} \eqdef \{\xv \in \F_2^n : \tau_h(\xv,\vec{0}) = a\}$. We are interested in the following quantity\footnote{The number of arguments of $A(\cdot)$ will make it clear whether we talk about bounds on general codes or on constant-weight codes.}, 
$$ 
A(n,d,a) \eqdef \max\left\{|\C| : \; \C \subseteq \mathcal{S}_{a}^{n,2}, \ d^{\textup{H}}_{\textup{min}}(\C) = d\right\}.
$$
Again, we define the asymptotic rate of constant-weight codes of relative weight $\alpha$ and with a certain relative minimum distance~$\delta$, 
$$ 
R(\delta,\alpha) = \mathop{\overline{\lim}}\limits_{n \to +\infty} \frac{1}{n} \log_{2}A(n,\lfloor \delta n\rfloor,\lfloor \alpha n \rfloor).
$$
Studying this quantity is interesting for its own sake but it is of additional importance because it can be used to obtain bounds on $R(\delta)$ thanks to the Elias-Bassalygo relation,
\begin{equation}\label{Eq:EB0}
	\forall \alpha \in \left[0,\frac{1}{2}\right], \quad R(\delta) \le 1 - h(\alpha) + R(\delta,\alpha)
	\end{equation}
	where $h(x) \eqdef - x \log_{2} x - (1-x)\log_{2}(1-x)$ denotes the binary entropy.

\subsection{An Overview of Some Different Bounds and Linear Programming Bounds}

The best lower  bound on $R(\delta)$ is the so-called Gilbert-Varshamov bound~\cite{G52,V57},
$$ R(\delta) \ge  R_{\textup{GV}}(\delta) \eqdef 1 - h(\delta).$$
It turns out that this bound corresponds to the minimum distance that is obtained by choosing a random linear code of the appropriate size. We expect this bound to be tight even though in the $q$-ary setting (when working in $\mathbb{F}_{q}^{n}$), there are codes that have a better minimum distance than random linear codes of the same size as soon as $q = p^{2}$ with $p \geq 7$ \cite{TVZ82} so the whole picture is not entirely clear.

On the other hand, there are various upper bounds on $R(\delta)$. The simplest known upper bound is the combinatorial Hamming bound. This bound was improved independently by Elias (attributed to Elias in \cite{SGB67}) and Bassalygo \cite{Bas65} also by using combinatorial arguments. These bounds are the following,
\begin{align*}
R(\delta) & \le R_{\textup{Hamm}}(\delta) \eqdef 1 - h\left(\frac{\delta}{2}\right), \\
R(\delta) & \le R_{\textup{EB}}(\delta) \eqdef 1 - h\left(\frac{1}{2} - \frac{1}{2}\sqrt{1 - 2\delta}\right).
\end{align*}

The question of finding codes with minimum distance $d$ can be generalized to the following question: given a set $\X$ and some distance function $\tau$ over $\X$, what is the maximum size of a subset $\C \subseteq \X$ such that each pair of distinct point of $\C$ have distance at least $d$? Delsarte introduced the important notion of association schemes~\cite{Del72,D73} that can in particular help solving this question. More precisely, if $(\X,\tau)$ satisfies certain conditions, then using the theory of association schemes, Delsarte shows how to construct a linear program such that its maximum value will be a bound on the maximum size of $\C$ with minimum distance at least $d$. An overview of Delsarte's theory can be found in~\cite{DL98}.

\begin{figure}[!ht]
	\includegraphics[height = 7cm ]{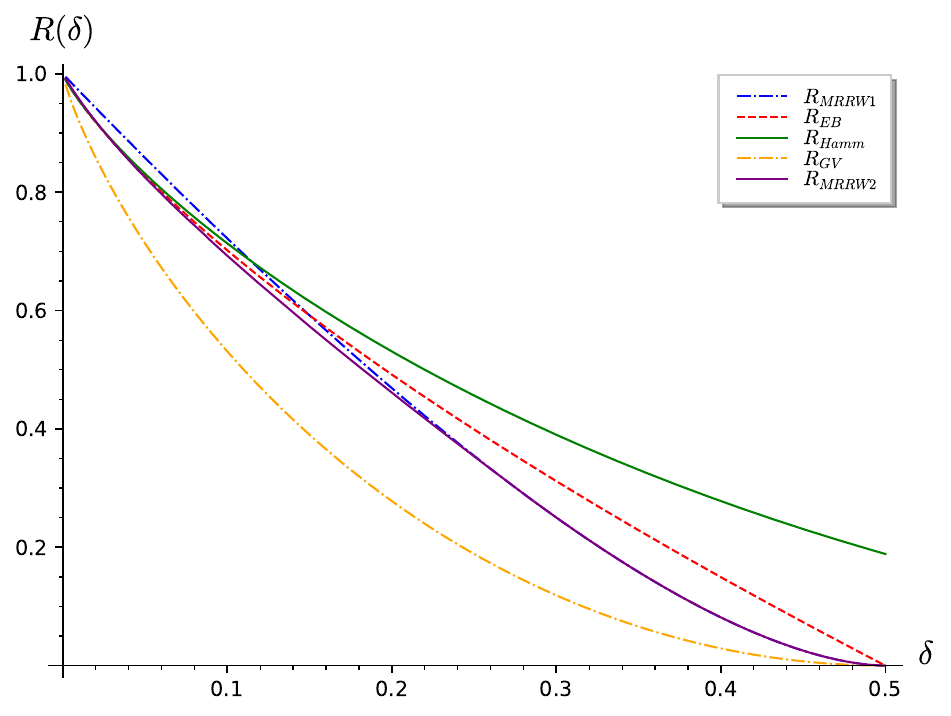}

	\caption{Known upper bounds and the Gilbert-Varshamov lower  bound on the asymptotic rate of binary codes $R(\delta)$ as function of their relative minimum distance~$\delta$.}
	\label{Figure:1}
\end{figure}

When instantiated in the boolean cube, this linear program involves Krawtchouk polynomials and MacWilliams identities. Its maximum objective $A_{\textup{LP}}(n,d)$ satisfies $A(n,d) \le A_{\textup{LP}}(n,d)$. Then, it is possible to deduce upper bounds on $A_{\textup{LP}}(n,d)$ by finding solutions to the associated dual linear program. Again, as we are interested in asymptotic upper bounds, we define, 
$$
R_{\textup{LP}}(\delta) \eqdef \mathop{\overline{\lim}}\limits_{n \to +\infty} \frac{1}{n} \log_2 A_{\textup{LP}}(n,\lfloor \delta n \rfloor).
$$ 
Using Delsarte's approach, McEliece, Rodemich, Rumsey and Welch \cite{MRRW77} proved what is now called the first linear programming bound,
$$ 
R(\delta) \le R_{\textup{LP}}(\delta) \le R_{\textup{MRRW1}}(\delta) \eqdef h\left(\frac{1}{2} - \frac{1}{2}\sqrt{\delta(1-\delta)}\right).
$$
It turns out that Delsarte's linear program approach can also be used to obtain bounds on the size of constant-weight codes with a certain minimum distance. It yields a linear program involving dual Hahn polynomials such that its optimum $A_{\textup{LP}}(n,d,a)$ satisfies $A(n,d,a) \le A_{\textup{LP}}\left(n,\lfloor \frac{d}{2} \rfloor,a\right)$\footnote{The factor two in the distance comes from the fact that the corresponding association scheme uses as its distance function $\tau_{\textup{H}}/2$.}. We denote the asymptotic value of this linear program,
$$
R_{\textup{LP}}(\delta,\alpha) =  \mathop{\overline{\lim}}\limits_{n \rightarrow \infty} \frac{1}{n} \log_2 A_{\textup{LP}}(n,\lfloor \delta n \rfloor,\lfloor \alpha n \rfloor).
$$ 
Again, by finding solutions to the associated dual linear program, it was shown in~\cite{MRRW77} that,
\begin{align}\label{Eq:R'}
R(\alpha,\delta) \le R_{\textup{LP}}\left(\alpha,\frac{\delta}{2}\right) \le R_{\textup{MRRW}}(\alpha,\delta) \eqdef   \frac{1}{2}\left( 1 - \sqrt{1-4\left(\sqrt{\alpha(1-\alpha)-\delta(1-\delta)}-\delta \right)^{2} }\right).
\end{align}
This bound is of particular importance because it can be combined with the Elias-Bassalygo relation (Equation \eqref{Eq:EB0}) to prove the so-called second linear programming bound, 
\begin{align*}
R(\delta) \le  R_{\textup{MRRW2}}(\delta) \eqdef \max_{0 \le \alpha \le \frac{1}{2}}\left\{1 - h(\alpha) + R_{\textup{MRRW}}(\alpha,\delta)\right\}
\end{align*}
This bound is, for the last 47 years, the best known upper bound on $R(\delta)$ for any $\delta$, but it is quite far from the lower  bound $R_{\textup{GV}}(\delta)$.  Note that even if there were some improvements on~$R(\delta,\alpha)$ for some parameters (see \cite{ABL99,Sam01} for example), these do not yield any improvements on $R(\delta)$ by using Equation~\eqref{Eq:EB0}. Surprisingly, Rodemich proved~\cite{Rod80} (see also~\cite{Del94,AB99,Sam01}) a lifting theorem which shows how to construct solutions to Delsarte's dual linear program on the boolean cube from a solution to the dual linear program for constant-weight codes. In particular, Rodemich has shown
how to use Equation~\eqref{Eq:R'} to prove, 
$$
R_{\textup{LP}}(\delta) \le R_{\textup{MRRW2}}(\delta)
$$
without using the Elias-Bassalygo relation. In other words, the best (current) solution of the linear program in the boolean cube leads to the second linear programming bound. It turns that this situation also appeared in the case of sphere packing in $\mathbb{R}^{n}$. An equivalent of the second linear programming bound from \cite{MRRW77} has been provided in \cite{KL78} via a dual linear program from spherical codes \cite[Ch. 9]{CS88} combined with an analogue of the Elias-Bassalygo relation \cite[Ch 9, Th. 6]{CS88}. The underlying bound is the current best upper bound for sphere packing in all sufficiently high dimensions. In \cite{CZ12}, a translation of Rodemich's theorem to $\mathbb{R}^{n}$ was used to provide a solution to the equivalent of the dual Delsarte's linear program in~$\mathbb{R}^{n}$ leading, as in the code-case, to the best known solution. An overview of these different bounds in the code-case on~$R(\delta)$ is depicted in Figure~\ref{Figure:1}.

While \cite{MRRW77} restricts its bounds to the binary case, an interesting question is also to provide bounds on the maximal size $A^{(q)}(n,d)$ of a code in~$\F_q^n$ of minimum distance $d$ and on the asymptotic rate $R^{(q)}(\delta) = \mathop{\overline{\lim}}\limits_{n \to +\infty} \frac{1}{n}\log_q A^{(q)}(n,\lfloor \delta n \rfloor)$ of codes in~$\F_q^n$, with~$q>2$ a prime power, as function of their relative minimum distance~$\delta$.
The first linear programming bound can easily be extended to the $q$-ary case \cite[\S F.]{DL98},
\begin{align*}
	R^{(q)}(\delta) & \le R^{(q)}_{\textup{LP}}(\delta) \le R^{(q)}_{\textup{MRRW1}}(\delta) \eqdef h_{q}\left( \gamma_{q}(\delta)\right) \mbox{ where } \\
	h_{q}(x) \eqdef -(1-x)\log_{q}(1-x) - &x\log_{q}\left(\frac{x}{q-1}\right) \quad , \quad \gamma_{q}(\delta) \eqdef \frac{1}{q}\left( q-1 - (q-2)\delta -2\sqrt{(q-1)\delta(1-\delta)} \right).
\end{align*}
Constant weight codes on $\F_q^n$ do yield a bivariate association scheme \cite{TAG85,CVZZ24} and it is possible to give a second linear programming bound in this setting. However, it is unclear whether Rodemich's lifting theorem can be applied to use these bounds in order to provide a bound directly on~$R^{(q)}_{\textup{LP}}(\delta)$.

Furthermore, Hamming and Elias-Bassalygo bounds are also known in this setting by using the same combinatorial arguments. For instance, the asymptotic Elias-Bassalygo bound is given in the $q$-ary case by, 
$$R^{(q)}(\delta) \le R^{(q)}_{\textup{LP}}(\delta)  \le R^{(q)}_{\textup{EB}} \eqdef 1 - h_{q}\left( J_{q}(\delta)\right) \quad \mbox{ where } \; J_{q}(\delta) \eqdef \left( 1 - \frac{1}{q} \right)\cdot \left( 1 - \sqrt{1-\frac{q\delta}{(q-1)}}\right).$$
The Elias-Bassalygo bound via the linear program method for the $q$-ary hypercube also exists and was proven in Aaltonen's thesis~\cite{Aal81}, where he also mentions the possibility to extend this to constant weight codes.

\subsection{Understanding the Limits of the Linear Programming Approach}
Delsarte's linear program approach is currently the most efficient one to provide bounds on $R(\delta)$. A natural question is therefore whether the second linear programming bound is the best one that can be achieved with this method. Also, there has recently been proposals for a hierarchy of linear programs involving  a generalized Delsarte's linear program on the boolean cube whose objective function is exactly $A(n,d)$ when going far enough in the hierarchy \cite{CJJ22,LL23a,CJJ23}. Unfortunately, these linear programs are currently too complicated to find new bounds and understanding solutions of Delsarte's linear program could be useful for finding new solutions to these more general linear programs. 

More concretely, what do we know about $R_{\textup{LP}}(\delta)$? The best upper bound is~$R_{\textup{LP}}(\delta) \le R_{\textup{MRRW2}}(\delta)$ using the MRRW bound for constant-weight codes and Rodemich's lifting theorem. On the other hand, regarding lower  bounds, the following bounds are known, which were proven respectively by Samorodnitsky~\cite{Sam01} and by Navon and Samorodnitsky~\cite{NS05}, 
\begin{align*}
	R_{\textup{LP}}(\delta)  & \ge R_{\textup{LP}}^{\textup{LWB1}}\eqdef \frac{1}{2}\left(R_{\textup{GV}}(\delta) + R_{\textup{MRRW1}}(\delta)\right), \\
	R_{\textup{LP}}(\delta) & \ge R_{\textup{LP}}^{\textup{LWB2}} \eqdef \frac{1}{2}h\left(1 - 2\sqrt{\delta(1-\delta)}\right).
\end{align*}

\begin{figure}[!ht]
	\includegraphics[height = 7cm]{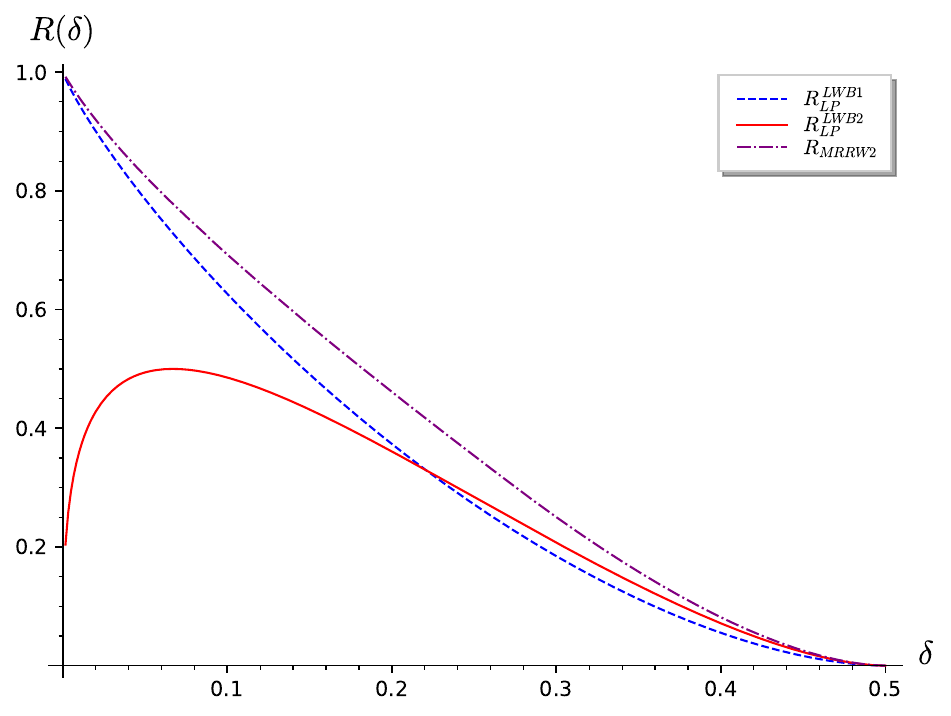}
	\caption{Best upper bound $R_{\textup{MRRW2}}(\delta)$ on Delsarte's linear program instantiated in the boolean cube and known lower  bounds on this program as function of the relative minimum distance $\delta$.}
	\label{Fig:2}
\end{figure}

In the attempt to understand the tightness of $R_{\textup{LP}}^{\textup{LWB1}}$, Barg and Jaffe~\cite{BJ99} performed numerical simulations for $A_{\textup{LP}}(n,d)$ with $n = 1000$. Their numerical simulations have shown that $R_{\textup{LP}}(\delta)$ is actually likely to be close to $R_{\textup{MRRW2}}(\delta)$. The above two lower  bounds and this upper bound are depicted in Figure~\ref{Fig:2}.

\subsection{The Laplacian Technique}The linear programming technique seems to have intrinsic limitations so it is important to consider other approaches. In this attempt, an extensive line of work started with the approach of Friedman and Tillich~\cite{FT05} which relied on graph theory and Fourier analysis over the Hamming cube. Roughly speaking, \cite{FT05} study the following: given a {\it linear} code $\mathcal{C}$, start from a random element of its dual and then walk on the Hamming cube 
according to some distribution $f$. Then, it turns out than an upper bound over the size of $\mathcal{C}$ can be found by using a function satisfying,
	\begin{align}\label{Eq:Lambda}
	\one_{\{1\}} \star f \ge \lambda f.
	\end{align}
	for some real $\lambda$, where $\one_{\{1\}}$ denotes the indicator function of words with Hamming weight $1$ and~$\star$ is the canonical convolution product. The value $\lambda$ is related to the minimum distance of $\C$ while the obtained upper bound on $\C$ can be expressed as a function of the range of $f$. In short, Friedman and Tillich find a function $f$ with small range such that Equation~\eqref{Eq:Lambda} is satisfied for a large $\lambda$. They use arguments from random walks and graph theory and recover the first linear programming bound with this approach seemingly orthogonal to Delsarte's linear program. Notice that the operation $\one_{\{1\}} \star f$ consists in performing an extra random step of Hamming distance $1$ after applying $f$. Because the operation $\one_{\{1\}} \star f$ is closely related to the Laplacian of $f$, we call this approach the Laplacian technique.

Several other works~\cite{Sam01,NS05} have followed the Laplacian technique by using Equation~\eqref{Eq:Lambda} in a crucial way. It has been shown among others that it can also be interpreted in terms of covering radius of the dual graph and can even be extended to non-linear codes~\cite{NS07}. 
	However, these methods are all tailored for the first linear programming by using Fourier analysis in the boolean cube. It is an open question (see~\cite{NS07} for instance) whether these methods can be adapted to the second linear programming bound.

\subsection{Contributions}

Our first contribution is to give generic solutions to Delsarte's dual linear programs. We show how to extend known solutions for the Hamming bound into solutions that achieve a generalized Elias-Bassalygo bound. This result is very general and can be applied to any $P$-polynomial association scheme under mild conditions. As it turned out, such solutions in the case of the $q$-ary Hamming scheme were already given by Aaltonenn~\cite{Aal81} and our results generalize this work to any association scheme constructed from a distance function. These solutiong can give better solutions to the linear program than MRRW-type solutions for small distances and show how to overcome some of the difficulties in improving the first linear programming bound~\cite{Sam23b}.

Our second contribution is to find explicit MRRW type solutions to Delsarte's linear program for essentially any $Q$-polynomial scheme. We rely on functions $f$ satisfying,
\begin{align}\label{Eq:fostar} \one_{\{1\}} \ostar f \ge \lambda f \end{align}
where $\ostar$ is not the canonical convolution of the association scheme derived from its underlying adjacency matrices and $P$-polynomials, but the convolution product derived from $Q$-polynomials. This can be seen as a direct ``dual'' generalization of Friedman and Tillich' approach. We then show that the function,
$$
g \eqdef \one_{\{1\}} \ostar f \ostar f - (\lambda -1) (f \ostar f)
$$
gives a solution to the dual linear program. This is a generalization of the observation of Samorodnitsky \cite{Sam23a}. We also give an explicit construction for ``good'' functions $f$ satisfying Equation \eqref{Eq:fostar}, that depends only on the $Q$-polynomials of the association scheme, generalizing~\cite{LL22}. 

These results use a somewhat different language but are strongly related to the results of Barg and Nogin~\cite{BN06}. They show how to recover MRRW-type bounds by finding the maximum eigenvalue (and corresponding eigenvector) of a certain operator, which is closely related to $f \mapsto \one_{\{1\}} \ostar f$. Their approach can be translated into constructing the function $g$ we described above. Interestingly, we also show how that our proof of the Elias-Bassalygo bound via the linear program also uses a function $f$ satisfying Equation~\eqref{Eq:fostar}, but without using the extra convolution used in the function~$g$. 

When correctly instantiated to Hamming spheres, we recover $R_{\textup{MRRW}}(\delta,\alpha)$ so this technique can be seen in some sense as a generalization of the Laplacian technique to the second linear programming bound. One has to be careful though, because we work with the convolution product in the $Q$-polynomial world. We somehow need to do the ``in reverse'' argument  of the Laplacian technique and use a ``dual'' Laplacian technique. While this is very well defined in the framework of association schemes, we don't know how to interpret the operation $\one_{\{1\}} \ostar f$  in terms of random walks on the associated inherited graph from spheres or in terms of a covering radius in the dual space. The fact that we require the $Q$-polynomial convolution for this generalization rather illustrates the difficulties in finding such interpretations. 

Because our results are meant to be as general as possible, we rely on the full machinery of association schemes. We therefore present an extensive introduction to Delsarte's theory of association schemes before proving our results.

 	\section{Notation and Preliminaries on Association Schemes}

{\bf \noindent Basic Notation.} The notation $x \eqdef y$ means that $x$ is being defined as equal to $y$. Given a set~$\mathcal{S}$, its indicator function will be denoted $\one_{\mathcal{S}}$. For a finite set $\mathcal{S}$, we will denote by $|\mathcal{S}|$ its cardinality. Let $\llbracket a,b \rrbracket$ be the set of integers $\{a,a+1,\dots,b\}$. We use the Kronecker delta notation $\delta_i^j = 1$ if $i = j$ and $\delta_i^j = 0$ otherwise.

Matrices are denoted in bold capital letters such as $\vec{A}$. For a finite set $\mathcal{S}$, let $\mathbb{C}(\mathcal{S}^{2})$ be the set of square matrices of order $|\mathcal{S}|$ and whose coefficients belong to $\mathbb{C}$. We will use the standard inner product on $\mathbb{C}\left( \mathcal{S}^{2} \right)$, \ie{} for $\vec{A},\vec{B} \in \mathbb{C}\left( \mathcal{S}^{2}\right)$,
$$
\braket{\vec{A}}{\vec{B}} \eqdef \tr\left( \vec{A}\vec{B}^{\dagger} \right) \quad , \quad \| \vec{A} \| \eqdef \sqrt{\braket{\vec{A}}{\vec{A}}}
$$ 
where $\vec{B}^{\dagger}$ denotes the conjugate transpose of $\vec{B}$. For any fixed order over $\mathcal{S}$ and $x,y \in \mathcal{S}$, we write~$\vec{A}(x,y)$ to denote the coefficient of $\vec{A}$ at row $x$ and column $y$. 
\medskip

Our aim now is to present needed  pre-requisites about association schemes. Almost all proofs are omitted. They can be found in the classical literature about association schemes like \cite{D73, BI84, DL98}. For the sake of completeness, we prove in Appendix \ref{app:proofs} all the claimed results.

\subsection{Equipartition Property and Association Schemes}\label{subsec:equiAssoc}

Let $\X$ be a finite set of ``points'' with~$|\X| \geq 2$ and let $\tau: \X^{2} \longrightarrow \llbracket 0,n \rrbracket$ be a distance function. Given~$(\X,\tau,n)$, we will consider the following adjacency matrices $\vec{D}_{i} \in \mathbb{C}(\X^{2})$ for $i \in \llbracket 0,n\rrbracket$, 
$$
\forall x,y \in \X, \quad \vec{D}_{i}(x,y) \eqdef \begin{cases}
1 \mbox{ if } \tau(x,y) = 1 \\
0 \mbox{ otherwise}
\end{cases}. 
$$    
Distance induced association schemes are triplets $(\X,\tau,n)$ satisfying the following properties.

\begin{restatable}[Equipartition Property and Non-Degenerate Triplets]{definition}{eqprop}\label{def:pijk}$(\X,\tau,n)$ is said to satisfy the equipartition property if for each $i, j, k \in\llbracket 0,n \rrbracket$, there exists a nonnegative integer $p^{k}_{i,j}$ such that,
	\begin{equation*}\label{eq:pijk} 
	\forall x,z \in \X \; \mbox{ such that } \; \tau(x,z) = k, \quad |\left\{ y \in \X: \; \tau(x,y) = i \;\mbox{ and }\; \tau(y,z) = j \right\}| = p^{k}_{i,j}.
	\end{equation*} 
	Furthermore, a triplet $(\X,\tau,n)$ satisfying the equipartition property is said to be non-degenerate if~$p_{1,k}^{k+1} \neq 0$ for all $k \in \llbracket 0,n-1 \rrbracket$. 
\end{restatable} 
The equipartition property ensures that the complex vector space generated by the adjacency matrices $\vec{D}_{i}$ is closed under matrix multiplication, \ie{} it forms an associative algebra. 

\begin{restatable}{proposition}{propoMultd}\label{propo:multD}
	Let $(\X,\tau,n)$ satisfying the equipartition property and let $\left( \vec{D}_{i} \right)_{i \in \llbracket 0,n \rrbracket}$ denote the associated adjacency matrices. We have,
	\begin{equation*}\label{eq:pijkDiDj}  
	\forall i,j \in \llbracket 0,n \rrbracket,\quad \vec{D}_{i}\cdot \vec{D}_{j} = \sum_{k\in \llbracket 0,n \rrbracket} p_{i,j}^{k} \vec{D}_{k} .
	\end{equation*} 
\end{restatable}
The above proposition is extremely powerful, it shows that the vector space generated by the $\vec{D}_{i}$ inherits a lot of structure. In particular,
by the symmetry of the distance $\tau$, the $p_{i,j}^{k}$ verify, 
\begin{equation*}
\forall i,j,k \in \llbracket 0,n \rrbracket, \quad p_{i,j}^{k} = p_{j,i}^{k}. 
\end{equation*} 
Therefore, it is readily seen 
that the complex vector space generated by the adjacency matrices~$\vec{D}_{i}$ is an associative algebra which is commutative. Furthermore, the fact that the $p_{i,j}^{k}$ are defined via the distance $\tau$ implies another strong property that will be useful:
\begin{equation}\label{eq:TriangleInequalityP}
	p_{i,j}^{k} = 0 \; \mbox{ if }\;  k > i+j,  \mbox{ or } \; |j-i| > k \; \mbox{ or as soon as} \; i,j,k > n .
\end{equation}

We are now ready to properly define distance induced association schemes\footnote{Distance-induced association schemes are also known as metric schemes.}.

\begin{restatable}{definition}{assocScheme}A distance induced association scheme is a triplet $(\X,\tau,n)$ where $\X$ is a finite set,~$\tau: \X^{2} \longrightarrow \llbracket 0,n \rrbracket$ is a distance and $(\X, \tau,n)$ satisfies the equipartition property and is non-degenerate.
\end{restatable}

\begin{remark}
	General association schemes are defined via a collection of some relations $(\mathcal{R}_{i})_{i\in \llbracket 0,n \rrbracket}$ over $\X^{2}$. Our definition restricts (which is enough for our purpose) to the case where the $\mathcal{R}_{i}$ are defined via $(x,y) \in \mathcal{R}_{i}$ if and only if $\tau(x,y) = i$. 
\end{remark}
\begin{remark}
	Association schemes that satisfy Equation \eqref{eq:TriangleInequalityP} are called $P$-polynomial association schemes. Any distance induced association scheme is $P$-polynomial. Conversely, for any $P$-polynomial association scheme, one can construct a distance function $\tau$ such that it is distance induced with respect to $\tau$~\cite[\S 5.2]{D73}. Also, $P$-polynomial schemes satisfy the following  $3$-term key relation,
	$$ 
	\Dm_1 \Dm_j = p_{1,j}^{j-1} \Dm_{j-1} + p_{1,j}^{j} \Dm_{j} + p_{1,j}^{j+1} \Dm_{j+1}.
	$$
\end{remark}

The most common association schemes (and the ones that we will consider) are,
\begin{itemize}
	\item {\bf The Hamming scheme:} $\X$ is the hypercube $\mathbb{F}_{q}^{n}$ endowed with Hamming metric,
	$$
	\forall \vec{x},\vec{y} \in \mathbb{F}_{q}^{n}, \quad \tau_{\textup{H}}(\vec{x},\vec{y}) \eqdef \left|\left\{ i \in \llbracket 1,n\rrbracket: \; x_{i} \neq y_{i} \right\}\right|.
	$$

	\item {\bf The Johnson scheme:} given some $a \in \llbracket 0,n \rrbracket$, $\X = \mathcal{S}_{a}^{n,2} = \{\xv \in \mathbb{F}_{2}^n : \tau_{\textup{H}}(\xv,\vec{0}) = a\}$ is the Hamming sphere of radius $a$ in the Hamming cube. The association scheme $(\X,\tau_{\textup{J}},a)$ is then defined with\footnote{One can check that for any $\xv,\yv \in \mathcal{S}_{a}^{n,2} $, $\tau_{\textup{H}}(\xv,\yv)$ is even hence $\tau_{\textup{J}}(\xv,\yv)$ is an integer.} $\tau_{\textup{J}} \eqdef \tau_{\textup{H}}/2$.  
\end{itemize}

\subsection{Fundamental Parameters of Association Schemes}

Let $(\X,\tau,n)$ be a distance induced association scheme. The corresponding matrices $\Dm_i$ are real and symmetric. Moreover, from Proposition~\ref{propo:multD} one can show that each $\Dm_i$ has $n+1$ distinct eigenvalues and that they share the same eigenspaces. Let $\Em_0,\dots,\Em_n$ be the projectors on these eigenspaces. 

One can prove that there exists an ordering on the matrices $(\Em_i)_{i \in \llbracket 0,n \rrbracket}$ such that, 
$$
\vec{E}_{0} = \frac{1}{|\X|}\sum_{i \in \llbracket 0,n \rrbracket} \vec{D}_{i}=  \frac{1}{{|\X|}}\cdot \vec{J} \quad \mbox{ where } \;  \vec{J} \; \mbox{ is the full-one matrix}.
$$

The fundamental parameters of an association scheme are defined with respect to an ordering of the matrices $\Em_0,\dots,\Em_n$. When we refer to an ordering $\Em_0,\dots,\Em_n$, we assume from now on that it satisfies $\Em_0 = \frac{1}{|\X|}\cdot \vec{J}$.

\begin{restatable}[$p$-numbers]{definition}{pnumbers}\label{def:pnumber} Let $(\X,\tau,n)$ be a distance induced association scheme with an ordering~$\Em_0,\dots,\Em_n$. Its underlying~$p$-numbers $p_{i}(j)$ are defined as,
\begin{equation*}
	\forall i \in \llbracket 0,n \rrbracket, \quad \vec{D}_{i} = \sum_{j\in \llbracket 0,n \rrbracket} p_{i}(j)\vec{E}_{j}.
\end{equation*}
\end{restatable}

Let us now introduce the norms (with a normalization) of these matrices $\vec{D}_{i}$ and $\vec{E}_{j}$.
\begin{restatable}{definition}{normvmi}
Let $(\X,\tau,n)$ be a distance induced association scheme with an ordering $\Em_0,\dots,\Em_n$, we define,
	$$
	\forall i \in \llbracket 0,n \rrbracket, \quad v_{i} \eqdef \frac{\| \vec{D}_{i}\|^{2}}{|\X|} \quad \mbox{and} \quad m_{i} \eqdef \| \vec{E}_{i} \|^{2} = \rank(\vec{E}_i).
	$$ 
\end{restatable}
The $v_{i}$ ({\it resp.} $m_{i}$) are called the valencies ({\it resp.} multiplicities) of the association scheme. From Definition \eqref{def:pijk}, one can obtain the following relation,
\begin{equation}\label{eq:vi} 
\forall i \in \llbracket 0,n \rrbracket, \quad v_{i} = p_{i,i}^{0}.
\end{equation} 
Matrices $(\vec{D}_{i})_{i \in \llbracket 0,n \rrbracket}$ and $(\vec{E}_{i})_{i \in \llbracket 0,n \rrbracket}$ generate the same Hilbert space which enables to define the $q$-numbers, an analogue of the $p$-numbers where the $\vec{E}_{i}$ and $\vec{D}_{i}$ are interchanged.

\begin{restatable}[$q$-numbers]{definition}{qnumbers}\label{def:qnumber} Let $(\X,\tau,n)$ be a distance induced association scheme with an ordering~$\Em_0,\dots,\Em_n$. Its underlying~$q$-numbers $q_{i}(j)$ are defined from the expansion of the orthogonal projectors $\vec{E}_{i}$ in the basis of adjacency matrices $(\vec{D}_{j})_{j \in \llbracket 0,n \rrbracket}$, \ie{} 
	\begin{equation*}
	\forall i \in \llbracket 0,n \rrbracket, \quad \vec{E}_{i} = \frac{1}{|\X|} \sum_{j \in \llbracket 0,n \rrbracket} q_{i}(j)\vec{D}_{j} .
	\end{equation*} 
\end{restatable}

Notice that the $p$ and $q$-numbers are real by symmetry of the $\vec{D}_{i}$, $\vec{E}_{j}$ and unicity of the decomposition in theses bases. Furthermore, from the definition of $\vec{E}_{0}$ as  $1/|\X|\sum_{i \in \llbracket 0,n \rrbracket} \vec{D}_{i}$ and the fact that $\vec{D}_{0} = \vec{Id} = \sum_{i \in \llbracket 0,n \rrbracket} \vec{E}_{i}$, we deduce that,
\begin{equation}\label{eq:q0j}  
\forall j \in \llbracket 0,n \rrbracket, \quad q_{0}(j) = 1 \quad , \quad p_{0}(j) = 1.
\end{equation} 

One can show that the matrices $(\Dm_i)_{i\in \llbracket 0,n \rrbracket}$ are pairwise orthogonal with respect to the inner product on matrices. Similarly the matrices $(\Em_i)_{i\in \llbracket 0,n \rrbracket}$ are pairwise orthogonal. This allows to show a strong relation between the $p$ and $q$-numbers.

\begin{restatable}{proposition}{proposcalProductEiDi}\label{propo:scalProductEiDi}
	Let $(\X,\tau,n)$ be a distance induced association scheme with an ordering~$\E_0,\dots,\E_n$.
$$
	\forall i,j \in \llbracket 0,n \rrbracket, \quad m_{j} {p_{i}(j)} = v_{i}q_{j}(i).
$$
\end{restatable}

The above relation is the key to prove many formulas involving $p$ and $q$-numbers. In particular, using that $q_{0}$ and $p_{0}$ are the constant functions equal to $1$ by Equation \eqref{eq:q0j}, we get $m_{0} = \| \vec{E}_{0} \|^{2} = 1$,~$v_{0} = \| \vec{D}_{0} \|^{2}/|\X| = 1$ and finally,
\begin{equation}\label{eq:p0j}
\forall i \in \llbracket 0,n \rrbracket, \quad p_{i}(0) = v_{i} \quad , \quad q_{i}(0) = m_{i}.
\end{equation}

\subsection{Algebra Structure for Pointwise Multiplication and $Q$-Polynomial Schemes}\label{subsec:pointWiseE}

We have deduced from Proposition \ref{propo:multD} that the vector space $\mathcal{H}$ generated by the $\vec{D}_{i}$ is closed for the standard matrix-product: it is an algebra. It is also closed under the pointwise multiplication~$(\vec{M},\vec{N}) \mapsto \vec{M} \circ \vec{N}$ be defined as,
$$
\vec{M} \circ \vec{N}(x,y) \eqdef \vec{M}(x,y)\vec{N}(x,y) .
$$  
Indeed the $\vec{D}_{i}$ verify $\vec{D}_{i} \circ \vec{D}_{j} = \delta_{i}^{j} \cdot \vec{D}_{i}$. But $\mathcal{H}$ is also generated by the $\vec{E}_{i}$ showing that we can 

define an equivalent of the $p_{i,j}^{k}$ (regarding Proposition \ref{propo:multD}): the $q_{i,j}^{k}$.

\begin{definition}\label{def:Krein}
	Let $(\X,\tau,n)$ be a distance induced association scheme with an ordering $\Em_0,\dots,\Em_n$. The underlying Krein parameters $q_{i,j}^{k}$ are defined from the expansion of $|\X|\cdot\vec{E}_{i} \circ \vec{E}_{j}$ in the basis~$(\vec{E}_{k})_{k\in \llbracket0,n\rrbracket}$, \ie{}
	$$
	\forall i,j \in \llbracket 0,n \rrbracket, \quad |\X| \cdot \vec{E}_{i}\circ \vec{E}_{j} = \sum_{k\in\llbracket 0,n \rrbracket} q_{i,j}^{k} \vec{E}_{k}. 
	$$
\end{definition}

The Krein parameters enjoy many interesting properties. For instance, they verify the following relations,
\begin{equation}\label{eq:Kreinqii0}
	\forall x \in \llbracket 0,n \rrbracket, \quad q_{x,x}^{0} = m_{x} > 0,
\end{equation}
\begin{equation}\label{eq:Kreinq10}
	\forall x \in \llbracket 0,n \rrbracket, \quad \sum_{y \in \llbracket 0,n \rrbracket} q_{y,1}^{x} = q_{1}(0),
\end{equation}
\begin{equation}\label{eq:invKrein}
	\forall x,y \in \llbracket 0,n \rrbracket, \quad m_{x} \cdot q_{y,1}^{x} = m_{y} \cdot q_{x,1}^{y}. 
\end{equation}
But most importantly, Krein parameters are nonnegative. 
\begin{restatable}{proposition}{positivityKrein}\label{propo:positivityKrein} $
	\forall i,j,k \in \llbracket 0,n \rrbracket, \quad q_{i,j}^{k} \geq 0 .
$
\end{restatable}

Krein parameters also appear when considering the product of $q$-numbers.

\begin{restatable}{proposition}{productKrein}\label{propo:productqnumbers}
	$
	\forall i,k,\ell \in \llbracket 0,n \rrbracket, \quad q_{k}(i)q_{\ell}(i) = \sum_{m\in \llbracket 0,n \rrbracket} q_{k,\ell}^{m} q_{m}(i).
$
\end{restatable}

Krein parameters $q_{i,j}^{k}$ are the dual of the $p_{i,j}^{k}$. However they are not in general integers. Furthermore, they don't necessarily verify the ``triangular inequality'' relation as given in Equation~\eqref{eq:TriangleInequalityP}. However there is a non-trivial (and important) subset of distance induced association schemes for which the Krein parameters verify this relation: $Q$-polynomial schemes.

\begin{definition}\label{def:Qpolynomial}
	A distance induced association scheme with an ordering $\Em_0,\dots,\Em_n$ is said to be~$Q$-polynomial if it satisfies the following two conditions,
	\begin{enumerate}
		\item $q_{i,j}^{k} = 0$ if $k > i+j$, or $|j-i| > k$ or as soon as $i,j,k > n$,

		\item $q_{1,k}^{k+1} \neq 0$ for all $k \in \llbracket 0,n \rrbracket$.
	\end{enumerate}
\end{definition}

This property implies in particular the $3$-term order relation,
$$ 
|\X|\cdot\Em_1 \circ \Em_j = q_{1,j}^{j-1} \Em_{j-1} + q_{1,j}^{j} \Em_{j} + q_{1,j}^{j+1} \Em_{j+1}
$$
which will be crucial in Subsection \ref{subsec:MRRWBounds} to recover linear programming bounds from~\cite{MRRW77}.

\subsection{Fourier Transforms and Convolutions}
Delsarte's theory of association schemes has strong relations with Fourier analysis via the use of Gelfand pairs~\cite{Del72,Dia88,CST08}. This creates a natural notion of ``Fourier transform'' and while we won't dive into this more general theory,  we introduce what we call the Fourier transform and its inverse (usually called $P$ and $Q$-transforms in the language of association schemes). All the definitions of this subsection are with respect to a fixed distance induced association scheme $(\X,\tau,n)$ with an ordering $\Em_0,\dots,\Em_n$.

\begin{definition}
	Given $f : \llbracket 0,n \rrbracket \longrightarrow \mathbb{C}$, we define its Fourier transform $\hf$ and its inverse Fourier transform $\tf$ as follows,
	\begin{align*}
	\hf(x) \eqdef \sum_{y\in \llbracket 0,n \rrbracket} f(y) p_y(x) \quad , \quad \tf(x) \eqdef  \frac{1}{|\X|}\sum_{y\in \llbracket 0,n \rrbracket} f(y) q_y(x).
	\end{align*}
\end{definition}
Simple examples of Fourier transform and its inverse are given by,
\begin{equation}
	\widehat{\one_{\{u\}}} = p_{u} \quad , \quad \widetilde{\one_{\{u\}}} = \frac{1}{|\X|}\;q_{u}. 
\end{equation}
When dealing with the Fourier transform and its inverse the following definition will be especially useful. 
\begin{definition}Given $f: \llbracket 0,n \rrbracket \longrightarrow \mathbb{C}$, we define its associated $\Dm$-matrix and $\Em$-matrix as follows,
	$$
	\vec{D}^{f} \eqdef \sum_{x \in \llbracket 0,n \rrbracket} f(x)\vec{D}_{x} \quad , \quad \vec{E}^{f} \eqdef \sum_{x \in \llbracket 0,n \rrbracket} f(x)\vec{E}_{x}.
	$$
\end{definition}

First notice that since $\D_i \circ \D_j = \delta_i^j \D_i$ and $\E_i \E_j = \delta_i^j \E_i$, we have,
$$ \D^f \circ \D^g = \D^{f\cdot g} \quad , \quad \E^f\cdot \E^g = \E^{f\cdot g}.$$
Moreover, notice that by the decompositions given in Definitions \ref{def:pnumber} and \ref{def:qnumber}, we have defined $\hf$ and $\tf$ to ensure~$\D^f = \E^{\hf}$ and $\E^f = \D^{\tf}$
which implies by unicity in the decomposition in bases $(\vec{D}_{i})_{i \in \llbracket 0,n \rrbracket}$ and $(\vec{E}_{i})_{i \in \llbracket 0,n \rrbracket}$ that,
 \begin{equation}\label{eq:hatTildef}
 	\widetilde{\widehat{f}} = \widehat{\widetilde{f}} = f.
 \end{equation}
 
 We can now define the convolution product and reverse convolution product between functions.
\begin{definition}
	Given $f,g : \llbracket 0,n \rrbracket \longrightarrow \mathbb{C}$, we define their convolution $\star$ and their reverse convolution~$\ostar$ as follows,
$$
	(f \star g)(x)  \eqdef \sum_{y,z \in \llbracket 0,n \rrbracket} f(y)g(z)p_{y,z}^x \quad , \quad 
	(f \ostar g)(x)  \eqdef \frac{1}{|\X|} \sum_{y,z\in \llbracket 0,n \rrbracket} f(y)g(z)q_{y,z}^x.
$$
\end{definition}
The convolution and reverse convolution are defined to ensure $\vec{D}^{f \star g} = \vec{D}^f \cdot \vec{D}^g$ and $\vec{E}^{f \ostar g} = \vec{E}^f \circ \vec{E}^g$  
which enables to prove the following proposition.

\begin{proposition}\label{propo:productConv}
	Let $f,g : \llbracket 0,n \rrbracket \longrightarrow \mathbb{C}$, we have,
	$$
	(1) \;\; \widehat{f \star g} = \hf\cdot \hg \quad , \quad (2) \;\; \widetilde{f \ostar g} = \tf\cdot \tg \quad , \quad (3) \;\; \widehat{(f g)} = \hf \ostar \hg \quad , \quad (4) \;\; \widetilde{(f g)} = \tf \star \tg.
	$$
\end{proposition}
\begin{proof}
	We write,
	\begin{enumerate}
		\item $\Em^{\widehat{f \star g}} =  \Dm^{f \star g} = \Dm^f\cdot \Dm^g = \Em^{\hf}\cdot \Em^{\hg} = \Em^{\hf\cdot \hg}$,
		\item $\Dm^{\widetilde{f \ostar g}} =  \E^{f \ostar g} = \E^f \circ \E^g = \Dm^{\tf} \circ \Dm^{\tg} = \Dm^{\tf\cdot \tg}$,
		\item Use $\widehat{(f \cdot g)} = \widehat{\left(\widetilde{\widehat{f}} \cdot \widetilde{\widehat{g}}\right)} \ \ $ and apply (2),
		\item Use $\widetilde{(f \cdot g)} = \widetilde{\left(\widehat{\widetilde{f}} \cdot \widehat{\widetilde{g}}\right)} \ \ $ and apply (1),
	\end{enumerate}
where in (1) and (2) we conclude by using the unicity in the decomposition in bases $(\vec{D}_{i})_{i \in \llbracket 0,n \rrbracket}$ and $(\vec{E}_{i})_{i \in \llbracket 0,n \rrbracket}$.
\end{proof}
Finally, the fact that the Krein parameters are nonnegative implies the following.
\begin{proposition}\label{propo:positivityOstar}
	Given $f,g : \llbracket 0,n \rrbracket \longrightarrow \mathbb{R}_{\geq 0}$, we have $f \ostar g \geq 0$. 
\end{proposition}

\subsection{Codes and Dual Distance Distribution}\label{subsec:code:DualWD}
Given a distance induced association scheme, our aim is to provide upper bounds on the size of codes with a fixed minimum distance. These objects are defined as follows.

\begin{definition}[Code, distance distribution and minimum distance]\label{def:code}
	Let $(\X,\tau,n)$ denote a distance induced association scheme. Any subset $\mathcal{C} \subseteq \X$ is called a code.

	Given a code $\mathcal{C}$, we define its distance distribution as,
	$$
	\forall t \in \llbracket 0,n \rrbracket, \quad a(t) \eqdef \frac{1}{|\mathcal{C}|} \cdot \left|\left\{ (c,c')\in \mathcal{C}^{2}: \; \tau(c,c') = t \right\}\right|.
	$$
	The minimum distance of $\mathcal{C}$ is then defined as,
	$$
	\dmin(\mathcal{C}) \eqdef \min \left\{ \tau(c,c'): \; c,c' \in \mathcal{C} \mbox{ and } c \neq c'  \right\} = \min\{t \in \llbracket 1,n \rrbracket : a(t) \neq 0\}.
	$$
\end{definition}
\begin{remark}
We have normalized the distance distribution of codes to ensure $a(0) = 1$.	
\end{remark}

In the remainder of this section, we will use Dirac's \emph{bra-ket} notation for linear algebra. Dirac's notation is borrowed from quantum computing~\cite{NC10} and even though our work is unrelated with quantum computing, this notation is in our opinion an especially elegant  way of presenting the results of this section, particularly the generalization of MacWilliams identities.
 
More concretely, let $\X = \{x_1,\dots,x_N\}$. From any $x_i \in \X$ we associate the column vector $\ket{x_i}$ whose~$ i^{th}$ entry is $1$ while the others are $0$. Then we write $\ket{v}$ for any vector of the complex-vector space generated by the $\ket{x_i}$ for $i \in \llbracket 1,N \rrbracket$. We write for example, 
$$
\ket{v} = \begin{pmatrix}
	v_{1} \\
	\vdots \\
	v_{N}
\end{pmatrix} = \sum_{i = 1}^N v_i \ket{x_i}.
$$
For a column vector $ \ket{v} = \begin{pmatrix}
v_{1} \\
\vdots \\
v_{N}
\end{pmatrix} $, we also define the line vector,
$$
\bra{v} \eqdef \begin{pmatrix}
	\overline{v_{1}} & \dots & \overline{v_{N}}
\end{pmatrix}.
$$
In particular, $\bra{x_i}$ is the line vector whose $i^{th}$ entry is $1$ while the others are $0$. With this notation, the canonical inner product between vectors $\braket{v}{w}$ is the multiplication $\bra{v} \cdot \ket{w}$.
Notice also that any rank one matrix of $\mathbb{C}(\X^{2})$ can now be written as $\ket{v} \cdot \bra{w}$ which we write $\ketbra{v}{w}$.

We now relate the distance distribution of a code $\C$ with the underlying $\vec{D}_{i}$ matrices of the association scheme.

\begin{definition}  Given a code $\mathcal{C}$, let, 
	\begin{equation*}
	\ket{\psi_{\mathcal{C}}} \eqdef \frac{1}{\sqrt{|\mathcal{C}|}}\; \sum_{c \in \mathcal{C}}\ket{c}.
	\end{equation*}
\end{definition}
This vector relates the distance distribution of a given code and underlying adjacency matrices of the association scheme. We have the following relation.
\begin{proposition} Let $(\X,\tau,n)$ be a distance induced association scheme, we have,
\begin{equation}\label{eq:aD}
\forall t \in \llbracket 0,n \rrbracket, \quad  a(t) = \bra{\psi_{\mathcal{C}}}\vec{D}_{t}\ket{\psi_{\mathcal{C}}}.
\end{equation} 
\end{proposition}
\begin{proof}
	With our notation, we have 
	$ \vec{D}_t = \sum_{x,x' : \tau(x,x') =t} \ketbra{x}{x'},$ which gives,
	\begin{align*}
	\bra{\psi_{\mathcal{C}}}\vec{D}_{t}\ket{\psi_{\mathcal{C}}} = \frac{1}{|\C|} \sum_{c,c' \in \C} \sum_{\substack{x,x' \in \X \\ \tau(x,x') = t}} \bra{c} \cdot \ketbra{x}{x'} \cdot \ket{c'} 
	 = \frac{1}{|\C|} \sum_{c,c' \in \C} \sum_{\substack{x \in \X \\ \tau(x,c') = t}} \braket{c}{x} 
	 =  \frac{1}{|\C|} \sum_{\substack{c,c' \in \C \\ \tau(c,c') = t}} 1 
	\end{align*}
	which concludes the proof by definition of $a(t)$. 
\end{proof}

The distance distribution of codes plays an important role in providing upper bounds on their size given their minimum distance, in particular their ``dual'' which is defined as follows.

\begin{definition}[Dual distance distribution]\label{def:dual}Let $(\X,\tau,n)$ denote a distance induced association scheme with an ordering $\Em_0,\dots,\Em_n$. Given a code $\mathcal{C}$, we define its dual distance distribution as, 
	\begin{equation*}\label{eq:dual} 
	a'(t) \eqdef\bra{\psi_{\mathcal{C}}}\vec{E}_{t}\ket{\psi_{\mathcal{C}}} 
	= \frac{1}{|\X|}\; \sum_{x\in \llbracket 0,n \rrbracket} \triple{\psi_\C}{q_t(x) \vec{D}_x}{\psi_\C} = \frac{1}{|\X|}\; \sum_{x \in \llbracket 0,n \rrbracket} q_t(x)a(x).
	\end{equation*} 
\end{definition}
Interestingly, the dual distance distribution turns out to be nonnegative, result which is known as a MacWilliams identity.

\begin{proposition}\label{theo:a'positive}
	Let $(\X,\tau,n)$ denote a distance induced association scheme with an ordering~$\Em_0,\dots,\Em_n$. Let $\C \subseteq \X$. Its dual distance distribution verifies,
	$$
	\forall t \in \llbracket 0,n \rrbracket, \quad a'(t) \geq 0.
	$$
\end{proposition}
\begin{proof}
	By definition the $\vec{E}_{t}$ are projectors. Therefore we can write each $\Em_t = \sum_{i} \ketbra{v_i^{t}}{v_{i}^{t}}$ for some rank $1$ projectors $\ketbra{v_i^t}{v_i^t}$ . Plugging this expression into the definition of $a'$ leads to, 
	$$
	a'(t) = \bra{\psi_{\mathcal{C}}}\vec{E}_{t}\ket{\psi_{\mathcal{C}}} = \bra{\psi_{\mathcal{C}}} \sum_{i}\ketbra{v_i^{t}}{v_{i}^{t}} \ket{\psi_{\mathcal{C}}} = \sum_{i} \left| \braket{\psi_{\mathcal{C}}}{v_{i}^{t}}\right|^{2} \geq 0
	$$
	which concludes the proof.
\end{proof}

\subsection{Delsarte's Linear Program of Association Schemes}\label{Section:LP}
Let $(\X,\tau,n)$ be a distance induced association scheme. In the attempt to provide bounds on the maximum size of a code $\C \subseteq \X$ with minimum distance at least $d$, Delsarte introduced \cite{D73} the following linear program
\begin{align*}
	\textbf{Delsarte's Linear} &  \textbf{ Program (DLP)} \nonumber \\
	\textrm{ maximize } \sum_{t \in \llbracket 0,n \rrbracket} u(t)  \\
	u(0) & = 1 \\
	u(t) & = 0 \textrm{ for } t \in \llbracket 1, d - 1\rrbracket  \\
	u(t) & \ge 0 \textrm{ for } t \in \llbracket d,n \rrbracket  \\
	\sum_{t\in \llbracket 0,n \rrbracket} u(t)q_i(t) & \ge 0 \ \textrm{for } i \in \llbracket 0,n \rrbracket. 
\end{align*}
\begin{definition}
	Let $(\X,\tau,n)$ be a distance induced association scheme with an ordering $\Em_0,\dots,\Em_n$ and $d \in \llbracket 0,n \rrbracket$. We define~$A_{\textup{LP}}(n,d)$ to be the maximum of the above linear program. 
\end{definition}

The following proposition justifies the introduction of Delsarte's linear program to give upper bounds on the size of a code given its minimum distance.

\begin{proposition}\label{propo:CodeDLP}
	Let $\C \subseteq \X$ be a code with minimum distance at least $d$. We have,
	$$
	|\C| \le A_{\textup{LP}}(n,d).
	$$ 
\end{proposition}
\begin{proof}
	It is a simple consequence of the fact that the distance distribution $a(t)$ of $\mathcal{C}$ verifies the condition of Delsarte's linear program, in particular the positivity of its dual distance distribution (see Definition \ref{def:dual}) given by Proposition \ref{theo:a'positive}. 
\end{proof}

The above proposition shows that solving Delsarte's linear program gives upper bounds for code sizes. Finding the value of this linear program is a hard problem. In order to find upper bounds on this linear program one has to look at the dual linear program which is a linear program such that its minimum will be larger than $A_{\LP}(n,d)$.

A simpler but essentially equivalent way of formulating the dual linear program is via the following proposition (see for instance \cite[III. B]{DL98}) which finds solutions to the dual linear program - and hence gives upper bounds on $A_{\LP}(n,d)$ - via the choice of some function.

\begin{proposition}\label{Proposition:Main}
	Let $d \in \llbracket 0,n \rrbracket$ and $f : \llbracket 0,n \rrbracket \longrightarrow \mathbb{R}$ be a function such that,
	$$ 
	\hf \ge 0 \quad , \quad \hf(0) > 0 \quad , \quad \forall x \ge d, \ f(x) \le 0.
	$$
	Then,
	$$
	A_{\textup{LP}}(n,d) \le |\X| \cdot \frac{f(0)}{\hf(0)}.
	$$
\end{proposition}
\begin{proof}
		Let $u$ be a function that satisfies the constraints of the linear program and $f$ that satisfies the requirements of the proposition. Let,
		$$
		\forall i \in \llbracket 0,n \rrbracket, \quad u'(i) \eqdef \sum_{t \in \llbracket 0,n \rrbracket} u(t)q_{i}(t) \ge 0.
		$$
		First,
		$$
		 \sum_{x\in \llbracket 0,n \rrbracket} u'(x)\hf(x) = \sum_{y \in \llbracket 0,n \rrbracket} \left( \sum_{x \in \llbracket 0,n \rrbracket} q_{x}(y)\widehat{f}(x) \right) u(y) = |\X| \cdot \sum_{y \in \llbracket 0,n \rrbracket} \widetilde{\widehat{f}}(y)u(y) = |\X| \cdot \sum_{ y \in \llbracket 0,n \rrbracket} f(y)u(y) 
		$$
where in the last equality we used Equation \eqref{eq:hatTildef}. 
		  Therefore, we write
$$ u'(0)\hf(0) \le \sum_{x\in \llbracket 0,n \rrbracket} u'(x)\hf(x) =   |\X | \cdot \sum_{x\in \llbracket 0,n \rrbracket} u(x)f(x) \le |\X|\cdot u(0)f(0) = |\X| \cdot f(0).$$
    In order to conclude, we just have to compute $u'(0) = \sum_{t\in \llbracket 0,n \rrbracket} u(t)q_0(t) = \sum_{t \in \llbracket 0,n \rrbracket} u(t)$ (see Equation \eqref{eq:q0j}). From there, 
    $$  
    \sum_{t\in \llbracket 0,n \rrbracket} u(t) \le |\X| \cdot \frac{f(0)}{\hf(0)}.
    $$
    It ends the proof since this is true for any solution $u$ of the linear program. 
\end{proof}
Let us stress that finding functions $f$ satisfying the above conditions corresponds to finding solutions to Delsarte's dual linear program. 	\section{Packing Bounds for Association Schemes}

The best asymptotic upper bounds on the size of $q$-ary and constant-weight codes for a fixed minimum distance, \ie{} packing bounds, were obtained in \cite{MRRW77} via Delsarte's Linear Program (DLP), in particular using Proposition \ref{Proposition:Main}. The functions that achieve the first and second linear programming bounds are rather involved and use the so-called Christoffel-Darboux formulas for orthogonal polynomials.

Here, we present three different families of functions satisfying the conditions of Proposition~\ref{Proposition:Main}. The first function that we use is very simple as it is just a convolution of bounded indicator functions but it recovers the well-known Hamming bound which holds in any distance induced association scheme. This solution used to recover the Hamming bound was already known since the work of Delsarte \cite[\S 4.3.3]{D73}.  We made the choice to present this function as our second family of functions has been deduced from this quite simple choice. It takes a similar function 
to which we add the coefficient $(q_1(x) - q_1(d))$. This is actually similar to the MRRW construction but here, it is $f$ that has bounded support\footnote{In the sense that the support is restricted in $\llbracket 0,r \rrbracket$ with $r$ significantly smaller than $n$.} while the MRRW functions have $\hf$ with bounded support. With these functions, we obtain a generalized Elias-Bassalygo bound. We are speaking here of a ``generalized'' Elias-Bassalygo bound as when instantiated to the Hamming association scheme we are precisely getting the bound known as Elias-Bassalygo.
Our generalized Elias-Bassalygo bound is very general, it only requires a distance induced association scheme which is not the case of~\cite{MRRW77}-like bounds (see also \cite{DL98}). Indeed, best known packing bounds obtained via DLP are also asking the association scheme to be $Q$-polynomial. 

This situation is well illustrated by our third choice of function which recovers \cite{MRRW77} bounds when instantiated to the Hamming and Johnson association schemes (our function slightly differs from the one in \cite{MRRW77}). Indeed, our third and ultimate function requires the underlying association schemes to be $Q$-polynomial to verify conditions of Proposition~\ref{Proposition:Main}.  These functions are close to the MRRW functions but are related to the Laplacian approach and makes the link between the linear programming approach and the (dual) Laplacian approach.

\subsection{Generalized Hamming Bound for the Linear Program}

In the following proposition we present the solution for Delsarte's dual linear program that recovers the Hamming bound, following for example~\cite{D73}.

\begin{proposition}[Generalized Hamming Bound for DLP]\label{theo:Hamming}
	Let $(\X,\tau,n)$ be a distance induced association scheme with an ordering $\Em_0,\dots,\Em_n$. For any $d \in \llbracket 1,n \rrbracket$, we have,
	$$ 
	A_{\textup{LP}}(n,d) \le \frac{|\X|}{\sum_{x = 0}^{\lfloor \frac{d - 1}{2}\rfloor} v_x}.
	$$
\end{proposition}
\begin{proof}
	The proof strategy will be to construct a good function $f$ which satisfies the requirements of Proposition~\ref{Proposition:Main}. We choose,
	$$
	f \eqdef \one_{\leq \left\lfloor \frac{d-1}{2} \right\rfloor } \star \one_{\leq \left\lfloor \frac{d-1}{2} \right\rfloor } \quad \mbox{ where } \; \one_{\leq \left\lfloor \frac{d-1}{2} \right\rfloor }  \eqdef \sum_{x = 0}^{\left\lfloor \frac{d-1}{2} \right\rfloor} \one_{\{x\}}.
	$$
	We have, 
	\begin{align*}
		f(x) & = \sum_{y,z\in \llbracket 0,n \rrbracket} \one_{\le \left\lfloor \frac{d-1}{2} \right\rfloor}(y)\one_{\le \left\lfloor \frac{d-1}{2} \right\rfloor}(z)p_{y,z}^x = \sum_{y,z = 0}^{ \left\lfloor \frac{d-1}{2} \right\rfloor} p_{y,z}^x.
	\end{align*}
	Let $h = \one_{\leq \left\lfloor \frac{d-1}{2} \right\rfloor}$. Since $f = h \star h$, we have $\hf = (\widehat{h})^2$ which gives, 
	\begin{align*}
		\hf(x) & = \left(\sum_{y \in \llbracket 0,n \rrbracket} \one_{\le \left\lfloor \frac{d-1}{2} \right\rfloor}(y) p_y(x)\right)^2 = \left(\sum_{y = 0}^{ \left\lfloor \frac{d-1}{2} \right\rfloor} p_y(x)\right)^2.
	\end{align*}
	We clearly have $\hf \ge 0$ and $\hf(0) > 0$. Also, using Equation \eqref{eq:TriangleInequalityP}, we have $f(x) = 0$ when~$x \ge d \ge 2  \left\lfloor \frac{d-1}{2} \right\rfloor$. This means the conditions of Proposition \ref{Proposition:Main} are satisfied. We now compute by Equations~\eqref{eq:vi} and~\eqref{eq:p0j}, 
	\begin{align*}
		f(0) & =  \sum_{y,z = 0}^{ \left\lfloor \frac{d-1}{2} \right\rfloor} p_{y,z}^0 = \sum_{y = 0}^{ \left\lfloor \frac{d-1}{2} \right\rfloor} v_y, \\
		\hf(0) & = \left(\sum_{y = 0}^{ \left\lfloor \frac{d-1}{2} \right\rfloor} p_y(0)\right)^2 = \left(\sum_{y = 0}^{ \left\lfloor \frac{d-1}{2} \right\rfloor} v_y \right)^2.
	\end{align*}
We can now use Proposition \ref{Proposition:Main}, 
$$ 
A_{\textup{LP}}(n,d) \le |\X| \cdot \frac{f(0)}{\hf(0)} = \frac{|\X|}{ \sum_{y = 0}^{ \left\lfloor \frac{d-1}{2} \right\rfloor} v_y}
$$
which concludes the proof. 
\end{proof}

\subsection{Generalized Elias-Bassalygo Bound for the Linear Program}\label{subsec:EB} 
Here, we start again from a function $f = \one_u \star \one_u \ge 0$ and we do the following changes: we will take $u$ which is a little bit larger than $\lfloor \frac{d-1}{2} \rfloor$. This seems problematic since the function will be nonnegative for values above~$d$. To circumvent this, we also multiply this function by the term $(q_1(x) - q_1(d))$. This will ensure that the function $f$ has the good sign conditions and we show that it is possible to choose $u$ above  $\lfloor \frac{d-1}{2} \rfloor$ while at the same time preserving the positivity of $\hf$.

\begin{theorem}[Generalized Elias-Bassalygo Bound for DLP]\label{theo:Johnson} 
	
	Let $(\X,\tau,n)$ be a distance induced association scheme with an ordering $\E_0,\dots,\E_n$ and $d \in \llbracket 1,n \rrbracket$. Suppose that $q_1$ is decreasing. Let,
\begin{equation}\label{cdt:uEB}
	u \in \left\{u_{0} \in \llbracket 0,n \rrbracket: \; \frac{q_{1}(u_0)^{2}}{q_{1}(0)} \geq q_{1}(d)+1  \right\}
\end{equation} 
	Then,
	$$ 
	A_{\textup{LP}}(n,d) \le (q_1(0) - q_1(d)) \cdot \frac{|\X|}{v_{u}}.
	$$  
\end{theorem}

\begin{proof}
	Again, our strategy is to find a good function to use in Proposition \ref{Proposition:Main}. The function~$f$ that we use is,
	$$ 
	f(x) \eqdef (q_1(x) - q_1(d))\cdot (\one_{\{u\}} \star \one_{\{u\}})(x).
	$$
	As $q_{1}$ is decreasing and $\one_{\{u\}} \star \one_{\{u\}} \geq 0$, we have that $f(x) \leq 0$ for $x \geq d$.

	Let us compute $\widehat{f}$.  Let $h = \one_u \star \one_u$. We compute by using Proposition~\ref{propo:productConv},
	$$ 
	\widehat{h} = \left(\widehat{\one_{\{u\}}}\right)^2 = p_u^2 \quad , \quad \widehat{q_1} = |\X|\cdot \one_{\{1\}}.
	$$
	We have $f = q_1 h - q_1(d) h$ so using again Proposition~\ref{propo:productConv}, we obtain,
	$$ \hf = \widehat{q_1} \ostar \widehat{h} - q_1(d) \widehat{h},$$
	from which we deduce,
	\begin{equation}\label{eq:hFJohnson} 
	\hf = |\X| \cdot \one_{\{1\}} \ostar p_u^2 - q_1(d) \cdot p_u^2.
	\end{equation}
	Let us admit for now that, 
	\begin{equation}\label{eq:toProve}
		\hf \geq p_{u}^{2}\geq 0 
	\end{equation} 
		which enables to apply Proposition \ref{Proposition:Main}.

	We can now compute (with Equations~\eqref{eq:vi} and~\eqref{eq:p0j}),
	\begin{align*}
		f(0) &= (q_{1}(0)-q_{1}(d))\cdot  (\one_{\{u\}} \star \one_{\{u\}})(0) = (q_{1}(0)-q_{1}(d))\cdot  p_{u,u}^0 = (q_{1}(0)-q_{1}(d))\cdot v_u, \\
		\hf(0) & \ge p_u(0)^2 = v_u^2 > 0.
	\end{align*}
	Plugging this into Proposition \ref{Proposition:Main}, we obtain,
	$$ 
	A_{\textup{LP}}(n,d) \le |\X| \cdot \frac{f(0)}{\hf(0)} \le \left(q_{1}(0)-q_{1}(d)\right)\cdot \frac{|\X|}{v_u}.
	$$
	To conclude it remains to prove Equation \eqref{eq:toProve}. To this aim, according to Equation \eqref{eq:hFJohnson}, let us give a lower bound on $\one_{\{1\}} \ostar p_u^2$. We have the following computation,
		\begin{align}
		\one_{\{1\}} \ostar p_{u}^{2}(x) &=\frac{1}{|\X|}  \sum_{y \in \llbracket 0,n \rrbracket} p_{u}^{2}(y)q_{y,1}^{x} \nonumber \\
		&\geq \frac{1}{|\X|}  \frac{1}{\sum_{y \in \llbracket 0,n \rrbracket} q_{y,1}^{x}} \; \left( \sum_{y \in \llbracket 0,n \rrbracket} p_{u}(y)q_{y,1}^{x} \right)^{2} \qquad \left(\mbox{By convexity of $x \mapsto x^{2}$}\right) \nonumber \\
		&=\frac{|\X|}{q_{1}(0)} \; \left( \one_{\{1\}} \ostar p_{u}(x) \right)^{2} \label{eq:1ostarpu}  
		\end{align} 
		where in the last equality we used Equation \eqref{eq:Kreinq10}. Now we write,
		$$
		\vec{E}^{\one_{\{1\}} \ostar p_u} = \vec{E}_1 \circ \vec{E}^{p_u} = \vec{E}_1 \circ \vec{D}_u = \frac{1}{|\X|}\left(\sum_{j \in \llbracket 0, n \rrbracket} q_1(j) \vec{D}_j\right) \circ \vec{D}_u = \frac{q_1(u)}{|\X|} \vec{D}_u =  \frac{q_1(u)}{|\X|}\vec{E}^{p_u}
		$$
		from which we obtain $\one_{\{1\}} \ostar p_u = \frac{q_1(u)}{|\X|}p_u$.
 Plugging this into Equation~\eqref{eq:1ostarpu} leads to, 
		$$
		\one_{\{1\}} \ostar p_{u}^{2}(x) \geq \frac{q_{1}(u)^{2} }{q_{1}(0) \; |\X|}\; p_{u}^2(x) \geq \frac{(q_{1}(d) + 1)}{|\X|}p_{u}^2(x)
		$$
		where in the inequality we used the assumption on $u$. Therefore, plugging this into Equation \eqref{eq:hFJohnson} gives,
		$$
		\widehat{f} \geq (q_{1}(d)+1)p_{u}^{2} - q_{1}(d)p_{u}^{2} = p_{u}^{2}
		$$
		which concludes the proof. 
	\end{proof}

\begin{remark}
	When instantiating the above theorem, we will choose $u$ satisfying Condition \eqref{cdt:uEB} but which maximizes $v_{u}$ in order to to get the best upper bound as possible. 
\end{remark}

\subsection{The Dual Laplacian Argument}
Here, we give general statements showing how to use a function $f$ satisfying, 
$$ \one_{\{1\}} \ostar \widehat{f} \ge \lambda \widehat{f}$$
and certain properties 
to find functions satisfying the requirements of Proposition~\ref{Proposition:Main}. The proof of the generalized Elias-Bassalygo bound is implicitly using this approach via the following claim. 

\begin{proposition}
	Let $(\X,\tau,n)$ be a distance induced association scheme with an ordering~$\Em_0,\dots,\Em_n$. Let $d \in \llbracket 0,n \rrbracket$ and let $f: \llbracket 0,n \rrbracket \longrightarrow \mathbb{R}$ such that,
	\begin{equation}
		|\X| \cdot \one_{\{1\}} \ostar \hf \geq (q_1(d) + 1) \hf \quad , \quad \hf \ge 0 \quad , \quad \hf(0) > 0 \quad , \quad f \ge 0
	\end{equation}
	with $q_1$ decreasing.
	Then we have, 
	$$
	A_{\textup{LP}}(n,d) \le \left(q_1(0) - q_1(d)\right) \cdot |\X|\cdot \frac{f(0)}{\hf(0)}.
	$$
\end{proposition}
\begin{proof}
	We take $g(x) \eqdef \left( q_{1}(x) - q_1(d)  \right)\cdot f$. We have $g(x) \le 0$ for $x \ge d$ since $q_1$ is decreasing and~$f \ge 0$. Furthermore, by assumption on $\widehat{f}$, 
	$$
	\hg = |\X| \cdot \one_{\{1\}} \ostar \hf - q_1(d) \hf \ge \hf \ge 0.
	$$
	This means $g$ satisfies the constraints of Proposition \ref{Proposition:Main}. We also have,~$g(0) = (q_1(0) - q_1(d))\cdot f(0)$ and $\hg(0) \ge \hf(0) > 0$. Therefore we obtain,
	$$ 
	A_{\textup{LP}}(n,d) \le |\X|\cdot \frac{g(0)}{\hg(0)} \le (q_1(0) - q_1(d)) \cdot |\X| \cdot \frac{f(0)}{\hf(0)}
	$$
	which concludes the proof. 
\end{proof}
Notice that in the previous proposition we asked the function $f$ to be nonnegative which is quite restrictive. Fortunately, we can apply the above strategy (in the choice of $f$)  by enforcing its square to appear in order to ensure the positivity. However, it is at the cost of an extra convolution on the Fourier transform but it preserves the ``eigenvalue property'', \ie{} $\one_{\{1\}} \ostar \widehat{f} \geq \lambda \cdot \widehat{f}$, and more importantly, it enables more functions.

\begin{proposition}\label{propo:eigevalueWithConv} 
	Let $(\X,\tau,n)$ be a distance induced association scheme with an ordering~$\Em_0,\dots,\Em_n$ with $q_1$ decreasing. Let $d \in \llbracket 0,n \rrbracket$ and let $f: \llbracket 0,n \rrbracket \longrightarrow \mathbb{R}$ such that,
	\begin{equation}
		|\X| \cdot \one_{\{1\}} \ostar \hf \geq (q_1(d) + 1) \hf \quad , \quad \hf \ge 0 \quad , \quad \hf(0) > 0.
	\end{equation}
	Then we have,
	$$
	A_{\textup{LP}}(n,d) \le \left(q_1(0) - q_1(d)\right) \cdot |\X|\cdot \frac{f^2(0)}{(\hf \ostar \hf)(0)}.
	$$
\end{proposition}
\begin{proof}
The key point is to observe that, 
	$$
	|\X| \cdot \one_{\{1\}} \ostar \hf \ostar \hf \ge (q_1(d) +1)(\hf \ostar \hf). 
	$$
	Indeed, let $h \eqdef |\X| \cdot \one_{\{1\}} \ostar \hf - (q_1(d) + 1)\hf$. Notice that by assumption $h \ge 0$. Therefore, 
	\begin{align*}
		|\X| \cdot \one_{\{1\}} \ostar \hf \ostar \hf -  (q_1(d) +1)(\hf \ostar \hf) = h \ostar \hf \ge 0
	\end{align*}
	by Proposition \ref{propo:positivityOstar} since both $h$ and $\hf$ are nonnegative. Recall now that by Proposition \ref{propo:productConv},
	$$
	\hf \ostar \hf = \widehat{(f^2)}.
	$$ 
	Also, $\widehat{(f^2)}(0) = (\hf \ostar \hf)(0) \ge \hf(0)^2 > 0$. This means that we have,
	$$ 
	|\X| \cdot \one_{\{1\}} \ostar \widehat{(f^2)} \ge (q_1(d) + 1)\widehat{(f^2)}  \quad , \quad \widehat{(f^2)} = \hf \ostar \hf \ge 0  \quad , \quad \widehat{(f^2)}(0) > 0 \quad , \quad f^2 \ge 0. 
	$$ 
We can therefore use the previous proposition with $f^2$. We obtain,
	$$
	A_{\textup{LP}}(n,d) \le \left(q_1(0) - q_1(d)\right) \cdot |\X|\cdot \frac{f^2(0)}{(\hf \ostar \hf)(0)}
	$$
	which concludes the proof.
\end{proof}

\subsection{MRRW Bounds Using the Laplacian Method}\label{subsec:MRRWBounds} The last proposition of the above subsection can be used to derive packing bounds which turn out to be known as {\it MRRW bounds}. Indeed, when instantiated to the Hamming and Johnson association schemes we exactly recover bounds from \cite{MRRW77}. 

\begin{theorem}[MRRW Bound for DLP]\label{theo:eigenvalueBound}
	Let $(\X,\tau,n)$ be a distance induced association scheme with an ordering $\Em_0,\dots,\Em_n$ which is also $Q$-polynomial and let $d \in \llbracket 0,n \rrbracket$. Furthermore, suppose that $q_1$ is decreasing. 
	Let $r^{\perp}$ be an integer in $\llbracket 0,n \rrbracket$ such that,  
$$
 q_{1}(d) + 1 \leq q_{1}(r^{\perp}).
	$$ 
	We suppose that there exists $x \in \llbracket 1, n \rrbracket$ such that $q_{x}(r^{\perp}) \leq 0$ and we define,
$$
	r \eqdef \min\left\{ x \in \llbracket 1,n \rrbracket: \; q_{x}(r^{\perp}) \leq 0 \right\} = \min\left\{ x \in \llbracket 1,n \rrbracket: \; p_{r^{\perp}}(x) \leq 0 \right\}.
	$$
Then, 
	$$
	A_{\textup{LP}}(n,d) \leq (q_{1}(0) - q_{1}(d)) \cdot \sum_{x \in \llbracket 0,r-1 \rrbracket} m_{x} .
	$$
\end{theorem}

To prove this theorem we will rely on the following function.

\begin{proposition}\label{propo:fEigenvalue}
	Let $(\X,\tau,n)$ be a distance induced association scheme with an ordering~$\Em_0,\dots,\Em_n$ which is also $Q$-polynomial. Let $f$ be the function such that,
	$$
	\forall x \in \llbracket 0,n \rrbracket, \quad \widehat{f}(x) \eqdef \left\{ 
	\begin{array}{cc}
	\frac{q_{x}\left(r^{\perp}\right)}{m_{x}} & \mbox{ if } x \in \llbracket 0,r-1 \rrbracket \\
	0 & \mbox{ otherwise}
	\end{array}
	\right.
	$$
	where $r$ and $r^{\perp}$ are defined as in Theorem \ref{theo:eigenvalueBound}. Then, 
	$$
	(i) \; \widehat{f}\geq 0, \quad (ii) \; \widehat{f}(0) > 0, \quad \mbox{ and } \quad (iii) \;\; |\X| \cdot \one_{\{1\}}\ostar \widehat{f} \geq q_{1}(r^{\perp})\cdot \widehat{f}.
	$$
\end{proposition}
\begin{proof}
	By assumption on $r$ we have for all $x \in \llbracket 0,r-1 \rrbracket$, $q_{x}(r^{\perp}) \geq 0$ showing that $\widehat{f}$ is nonnegative. Furthermore, $\widehat{f}(0) =q_{0}(r^{\perp})/m_{0} = 1$. 
	Let us now show that condition $(iii)$ holds. First,
	\begin{align*} 
	\forall x \in \llbracket 0,r-2 \rrbracket, \quad   (\one_{\{1\}} \ostar \widehat{f}) (x) 
	&= \frac{1}{|\X|} \sum_{y \in \llbracket 0,r-1 \rrbracket} \widehat{f}(y) q_{1,y}^{x} \nonumber \\
	& = \frac{1}{|\X|} \sum_{y \in \llbracket 0,r-1 \rrbracket} \frac{q_{1,y}^{x}}{m_y} q_y(r^\perp) \nonumber\\ 
	& = \frac{1}{|\X|} \sum_{y \in \llbracket 0,r-1 \rrbracket} \frac{q_{1,x}^{y}}{m_x} q_y(r^\perp) \quad \left(\mbox{By Equation \eqref{eq:invKrein}}\right) \nonumber \\ 
& = \frac{1}{|\X|}\frac{1}{m_x} q_1(r^\perp)q_x(r^\perp) \\
	&= \frac{q_1(r^\perp) \widehat{f}(x)}{|\X|}
	\end{align*} 
	where we used Proposition \ref{propo:productqnumbers} combined with the fact that  the scheme is supposed to be $Q$-polynomial (see Definition \ref{def:Qpolynomial}) and the $q_{i,j}^{k}$ are equal to $0$ if one $i,j,k$ is greater than the sum of the other two. Furthermore, using once again this assumption,
	\begin{align*}
		(\one_{\{1\}} \ostar \widehat{f}) (r-1) & = \frac{1}{|\X|} \left(q_{1,r-2}^{r-1}\widehat{f}(r-2) + q_{1,r-1}^{r-1}\widehat{f}(r-1) +  q_{1,r}^{r-1}\widehat{f}(r)\right) \\
		& = \frac{1}{|\X|} \left(\frac{q_{1,r-2}^{r-1}}{m_{r-2}}q_{r-2}(r^\perp) + \frac{q_{1,r-1}^{r-1}}{m_{r-1}}q_{r-1}(r^\perp) \right) \\
		& = \frac{1}{|\X|} \left(\frac{q_{1,r-1}^{r-2}}{m_{r-1}}q_{r-2}(r^\perp) + \frac{q_{1,r-1}^{r-1}}{m_{r-1}}q_{r-1}(r^\perp) \right) \quad \left(\mbox{By Equation \eqref{eq:invKrein}}\right) \\
		& \ge \frac{1}{|\X|\; m_{r-1}} \left({q_{1,r-1}^{r-2}} q_{r-2}(r^\perp) + {q_{1,r-1}^{r-1}}q_{r-1}(r^\perp) + {q_{1,r}^{r-1}}q_{r}(r^\perp)\right) \\
		& = \frac{1}{|\X|\; m_{r-1}} q_1(r^\perp)q_{r-1}(r^\perp) = \frac{q_1(r^\perp)\widehat{f}(r-1)}{|\X|}
	\end{align*}
where we used for the inequality $q_{1,r}^{r-1}q_r(r^\perp) \le 0$ coming from the definition of $r$ and the positivity of the $q_{i,j}^{k}$ (see Proposition \ref{propo:positivityKrein}). 
	Finally,
	\begin{align*}
	(\one_{\{1\}} \ostar \widehat{f}) (r) = \sum_{y \in \llbracket 0,r+1 \rrbracket} \widehat{f}(y)q_{1,y}^r = \widehat{f}(r-1)q_{1,r-1}^r \ge 0 
	\end{align*}
	and
	$
	\forall x > r, \ (\one_{\{1\}} \ostar \widehat{f}) (x) = 0.
$
	From there, we deduce that,
	\begin{align*}
	\forall x \in \llbracket r,n \rrbracket, \quad (\one_{\{1\}} \ostar \widehat{f}) (x) \ge 0 = \frac{q_1(r^\perp)}{|\X|}\widehat{f}(x)
	\end{align*}
	which concludes the proof.  
\end{proof}
\begin{proof}[Proof of Theorem \ref{theo:eigenvalueBound}] First, the equality when defining $r^{\perp}$ comes from Proposition~\ref{propo:scalProductEiDi}.
	Let us now take $f$ as defined in the above proposition. Recall that by assumption,
	$$
	q_{1}(d) + 1 \leq q_{1}(r^{\perp}).
	$$ 
	We can therefore apply Proposition \ref{propo:eigevalueWithConv} (here is used the assumption that $q_{1}$ is a decreasing function) with the above function $f$. We get,
	$$
	\widehat{f} \ostar \widehat{f}(0) = \frac{1}{|\X|} \; \sum_{x\in \llbracket 0,n \rrbracket}  \widehat{f}(x)^{2}q_{x,x}^{0} =  \frac{1}{|\X|} \; \sum_{x\in \llbracket 0,r-1 \rrbracket}  \widehat{f}(x)^{2}m_x
	$$
	where in the last equality we used Equation \eqref{eq:Kreinqii0}. Furthermore,
	\begin{align*}
	f^2(0)  = \widetilde{\widehat{f}}(0)^{2}
	&= \frac{1}{|\X|^2} \left(\sum_{y \in \llbracket 0,n \rrbracket } \widehat{f}(y)q_y(0) \right)^2 \\
	& = \frac{1}{|\X|^2} \left(\sum_{y \in \llbracket 0,r-1 \rrbracket } \widehat{f}(y)m_y \right)^2 \quad \left(\mbox{By Equation \eqref{eq:p0j}}\right)\\
	& \le \frac{1}{|\X|^2}\left(\sum_{y \in \llbracket 0,r-1 \rrbracket}m_y\right) \left(\sum_{y \in \llbracket 0,r-1 \rrbracket} \widehat{f}(y)^{2}m_y\right)
	\end{align*}
	where in the last inequality we used the Cauchy-Schwartz inequality. From there, we get by applying Proposition~\ref{propo:eigevalueWithConv},
	$$
	A_{\textup{LP}}(n,d) \leq(q_{1}(0)-q_{1}(d)) |\X| \frac{f^2(0)}{(\hf \ostar \hf)(0)}  \le  (q_{1}(0)-q_{1}(d)) \cdot \sum_{y \in \llbracket 0,r-1 \rrbracket} m_{y}
	$$ 
	which concludes the proof. 
\end{proof} 	\section{Applications: Packing Bounds for $q$-ary and Constant-Weight Binary Codes}

We are interested in this section to give upper bounds on the size of $q$-ary and constant-weight binary codes, \ie{} subsets of $\mathbb{F}_{q}^{n}$ and $\mathcal{S}_{a}^{n,2}$ (words of Hamming weight $a$ in $\mathbb{F}_{2}^{n}$), for a fixed minimum distance. Bounds will also be presented asymptotically in $n$ and in order to describe them compactly let us introduce some notation. We define $A^{(q)}(n,d)$ ({\it resp.} $A(n,d,a)$) to be the largest possible codes of $\mathbb{F}_{q}^{n}$ ({\it resp.} $\mathcal{S}_{a}^{n,2}$) with minimum {\it Hamming distance} at least $d$. Next we define,
$$
R^{(q)}(\delta) \eqdef \mathop{\overline{\lim}}\limits_{n \to +\infty} \frac{1}{n} \log_q A(n,\lfloor \delta n \rfloor) \quad \left(\mbox{\it resp.} \; R(\delta,\alpha) \eqdef \mathop{\overline{\lim}}\limits_{n \to +\infty} \frac{1}{n} \log_2A(n,\lfloor \delta n \rfloor, \lfloor \alpha n \rfloor ) \right)
$$
Upper bounds over $R^{(q)}(\delta)$ will involve the $q$-ary entropy, 
$$
h_{q}: x\in [0,1] \longmapsto -(1-x)\log_{q}(1-x) - x\log_{q}\left(\frac{x}{q-1}\right).
$$
This function gives the asymptotic behaviour of the binomial coefficients as shown in the following elementary lemma which will be at the core of all asymptotic results of this section. 

\begin{lemma}\label{lemma:binEntropy}
	Let $t\eqdef \lfloor \tau n \rfloor$, we have,
	$$
	\frac{1}{n} \log_{q}\binom{n}{t}(q-1)^{t} \mathop{=}\limits_{n \to +\infty} h_{q}(\tau) + o(1).
	$$
\end{lemma}

\subsection{Hypercube Case}

We instantiate in this subsection packing-bounds from the previous section in the Hamming scheme $(\mathbb{F}_{q}^{n},\tau_{\textup{H}},n)$ where $\tau_{\textup{H}}$ denotes the Hamming distance. This association scheme comes with a canonical ordering $\Em_0,\dots,\Em_n$. We give in what follows all the fundamental parameters of this association scheme with respect to this ordering as well as required properties to apply Theorems \ref{theo:Hamming}, \ref{theo:Johnson} and \ref{theo:eigenvalueBound}. We refer the reader to \cite{DL98}.

First, $(\mathbb{F}_{q}^{n},\tau_{\textup{H}},n)$ is a distance induced association scheme which is also $Q$-polynomial. 

Its valencies and multiplicities are given by,
\begin{equation}\label{eq:valencyMultiplicityHS} 
\forall i \in \llbracket 0,n \rrbracket, \quad v_{i} = m_{i} = \binom{n}{i}(q-1)^{i}.
\end{equation} 
The $p$ and $q$-numbers of the Hamming scheme involve {\it Krawtchouk polynomials} $K_{k}^{n,q}$ which are defined as follows, 
$$
\forall k \in \llbracket 0,n \rrbracket, \quad K_{k}^{n,q}(X) \eqdef \sum_{j \in \llbracket 0,k \rrbracket} (-1)^{j}(q-1)^{k-j} \binom{X}{j}\binom{n-X}{k-j}
$$
where $\binom{X}{i} \eqdef {X(X-1)\cdots (X-i+1)}/{i!}$. 
More precisely, $p$ and $q$-numbers are the integers given by the evaluation of the Krawtchouk polynomials over~$\llbracket 0,n \rrbracket$, \ie{}
\begin{equation*}
\forall i,k \in \llbracket 0,n \rrbracket, \quad q_{k}(i) = p_{k}(i) =  K_{k}^{n,q}(i).
\end{equation*}
Then it is readily seen that,
\begin{equation*}
\forall i \in \llbracket 0,n \rrbracket, \quad q_{1}(i) = (q-1)(n-i) - i = (q-1)n - qi
\end{equation*} 
which is a decreasing function as required in Theorems \ref{theo:Johnson} and \ref{theo:eigenvalueBound}. We denote $A^{(q)}_{\textup{LP}}(n,d)$ the maximum value of the associated linear program (as per Subsection~\ref{Section:LP}) and we define, 
$$
R^{(q)}_{\textup{LP}}(\delta) \eqdef \mathop{\overline{\lim}}\limits_{n \to +\infty} \frac{1}{n} \log_qA^{(q)}_{\textup{LP}}\left(n,\lfloor \delta n \rfloor\right).
$$
We immediately deduce from Proposition~\ref{propo:CodeDLP} that, 
$$
A^{(q)}(n,d) \le A^{(q)}_{\textup{LP}}(n,d) \quad , \quad R^{(q)}(\delta) \leq R_{\textup{LP}}^{(q)}(\delta).
$$

{\bf \noindent Hamming Bound.} By using the valencies of $(\mathbb{F}_{q}^{n},\tau_{\textup{H}},n)$ and Theorem \ref{theo:Hamming} we easily recover the Hamming bound. 
\begin{proposition}[Hamming Bound for $A^{(q)}_{\textup{LP}}(n,d)$]\label{theo:H}
	For any $q,n \ge 2$ and $d \ge 1$,
	$$
	A^{(q)}(n,d) \le A^{(q)}_{\textup{LP}}(n,d) \le \frac{q^n}{\sum_{x=0}^{\left\lfloor \frac{d-1}{2}\right\rfloor} v_x} =
		\frac{q^n}{\sum_{x=0}^{\left\lfloor \frac{d-1}{2}\right\rfloor} (q-1)^{x}\binom{n}{x}},
	$$
	which implies asymptotically,
	$$
	R^{(q)}(\delta) \le R^{(q)}_{\textup{LP}}(\delta) \leq 1 - h_{q}\left( \delta/2 \right).
	$$
\end{proposition}

{\bf \noindent Elias-Bassalygo Bound.} Let us now instantiate to the hypercube our generalized Elias-Bassalygo bound of Theorem \ref{theo:Johnson}. As we show we indeed recover the bound classically known as Elias-Bassalygo.

\begin{theorem}[Elias-Bassalygo Bound for $A^{(q)}_{\textup{LP}}(n,d)$]\label{theo:EB} 
	For any $q,n \ge 2$ and $d \in \llbracket 0,\lfloor n(q-1)/q \rfloor \rrbracket$, we have,
	$$
	A^{(q)}(n,d) \le A^{(q)}_{\textup{LP}}(n,d) \leq qd \cdot \frac{q^{n}}{\binom{n}{u}(q-1)^{u}}, \quad \textrm{where} \quad u \eqdef  \left\lfloor n\frac{q-1}{q} \cdot \left( 1-\sqrt{1-\frac{qd-1}{(q-1)n}}\right) \right\rfloor.$$

	It implies asymptotically for any $\delta \in [0,(q-1)/q]$,
	$$
	R^{(q)}(\delta) \le R^{(q)}_{\textup{LP}}(\delta) \leq 1 - h_{q}\left( J_{q}(\delta)\right), \quad \textrm{where} \quad J_{q}(\delta) \eqdef \frac{q-1}{q}\cdot \left( 1 - \sqrt{1-\frac{q\delta}{(q-1)}}\right).
	$$
\end{theorem}

\begin{proof}
	Our strategy is to apply Theorem \ref{theo:Johnson}. 
First $q_{1}$ is indeed a decreasing function. Now, let us compute,
	$$
	u \in \left\{ u_{0}\in \llbracket 0,n \rrbracket: \; \frac{q_{1}(u_{0})^{2}}{q_{1}(0)} \geq q_{1}(d)+1 \right\} \mbox{ which maximizes } \; \binom{n}{u}(q-1)^{u}.
	$$
	We have the following computation,
	$$
	q_{1}(0)\left(q_{1}(d)+1\right) = (q-1)n\left( (q-1)n - qd + 1 \right) = \left( (q-1)n\right)^{2}\left( 1 - \frac{qd-1}{(q-1)n} \right)
	$$
	Since $d \le n(q-1)/q$, the right hand side term is non negative and we therefore have
	$$
	q_{1}(u_{0})^{2} \geq q_{1}(0)\left( q_{1}(d)+1 \right) \Leftrightarrow (q-1)n - qu_{0} \geq (q-1)n\sqrt{1-\frac{qd-1}{(q-1)n}}
	$$
	showing that 
we have to choose $u$ smaller than,
	$$
	\left\lfloor n \; \frac{q-1}{q}\cdot \left( 1-\sqrt{1-\frac{qd-1}{(q-1)n}}\right) \right\rfloor .
	$$
	We can choose $u$ as above as $y \mapsto \binom{n}{y}(q-1)^{y}$ is an increasing function over $\left\llbracket 0, \left\lfloor n \; \frac{q-1}{q} \right\rfloor \right\rrbracket$. 
	Applying Theorem \ref{theo:Johnson}, we obtain,
	$$ A^{(q)}_{\textup{LP}}(n,d) \le (q_1(0) - q_1(d))\frac{q^n}{v_u} = qd \cdot \frac{q^{n}}{\binom{n}{u}(q-1)^{u}}.$$
The asymptotic result easily follows from Lemma \ref{lemma:binEntropy}. 
\end{proof}

{\bf \noindent MRRW1 Bound.} We now instantiate the bound from Theorem \ref{theo:eigenvalueBound} to the Hamming scheme. We recover the bound from \cite[Eq. (2.6)]{MRRW77} but in the $q$-ary setting (\cite{MRRW77} restricts to the case $q=2$ but it was generalized for example in~\cite{DL98}).

\begin{theorem}[MRRW1-type Bound for $A^{(q)}_{\textup{LP}}(n,d)$]\label{theo:MRRW1}
	Let integers $q,n \ge 2$ and $d \in \llbracket 1,\lfloor n(q-1)/q \rfloor \rrbracket$. Let $r = \lceil \zeta_{d-1} \rceil$, where $\zeta_{d-1}$ is the first zero of $K^{n,q}_{d-1}(X)$.
	Then, 
	$$
	A^{(q)}(n,d) \le A^{(q)}_{\textup{LP}}(n,d) \le  qd \cdot \sum_{x=0}^{r-1} (q-1)^{x}\binom{n}{x},
	$$
	which implies asymptotically for any $\delta \in [0,(q-1)/q]$,
	\begin{equation*} 
	R^{(q)}(\delta) \le R^{(q)}_{\textup{LP}}(\delta) \le h_{q}\left( \gamma_{q}(\delta)\right) \mbox{ where } \; \gamma_{q}(\delta) \eqdef \frac{1}{q}\left( q-1 - (q-2)\delta -2\sqrt{(q-1)\delta(1-\delta)} \right).
	\end{equation*} 
\end{theorem}

\begin{proof} Our goal is to apply Theorem~\ref{theo:eigenvalueBound}. First, notice that $q_1(d) + 1 \le q_1(d-1)$ so we choose $r^\perp = d-1$. Let $r = \lceil \zeta_{r^\perp}\rceil$ where $\zeta_{r^{\perp}}$ denotes the first zero of $K_{r^{\perp}}^{n,q}(X)$ in $[0,n]$. We know that the zeros of this polynomial are all in this interval and are simple. Furthermore, there is always an integer between any two consecutive zeros \cite{CS90}. Therefore, as $K_{r^{\perp}}^{n,q}(0) = v_{r^{\perp}} > 0$, we obtain by continuity, 
\begin{align*}
	q_r(r^\perp) &= \frac{m_r}{v_{r^\perp}} \; p_{r^\perp}(r) = \frac{m_r}{v_{r^\perp}} \; K^{n,q}_{r^\perp}(r) \le 0, \\
	\forall x \in {\llbracket}0,r-1 \rrbracket,  \ q_{x}({r^\perp}) &=  \frac{m_x}{v_{r^\perp}} \; p_{r^\perp}(x) =  \frac{m_x}{v_{r^\perp}}\; K^{n,q}_{r^\perp}(x)  > 0.
\end{align*}

This shows that $r^{\perp}$ and $r$ satisfy the conditions of Theorem \ref{theo:eigenvalueBound}, which gives,
	$$ 
	A^{(q)}_{\textup{LP}}(n,d) \le (q_1(0) - q_1(d)) \cdot \sum_{x = 0}^{r-1} m_r = qd \cdot \sum_{x = 0}^{r-1} (q-1)^x \binom{n}{x}.
	$$

We now prove the asymptotic	part of the proposition for a fixed $\delta \in (0,(q-1)/q)$. The asymptotic part of the proposition follows from the asymptotic expansion of~$\zeta_{x}$ (which denotes the first root of~$K_{x}^{n,q}(X)$). Indeed, given $x \in \llbracket 0,n \rrbracket$ such that $x/n \mathop{\longrightarrow}\limits_{n \to +\infty} \alpha \in [0,(q-1)/q]$, we have~\cite[\S IV. F]{DL98}, 
	$$
	\frac{\zeta_{x}}{n} \mathop{=}\limits_{n \to +\infty} \gamma_{q}(\alpha) + o(1).
	$$ 
	In our case, $r^{\perp}/n = (d-1)/n \mathop{\longrightarrow}\limits_{n \to +\infty} \delta \in [0,(q-1)/q]$ (where $d = \lfloor \delta n \rfloor$). Therefore~$r/n = \lceil \zeta_{r^\perp} \rceil/n \mathop{\longrightarrow}\limits_{n \to +\infty} \gamma_{q}(\delta)$. We can conclude using Lemma~\ref{lemma:binEntropy} that 
	$R^{(q)}_{\textup{LP}}(\delta) \leq h_q(\gamma_q(\delta))$.
\end{proof}

\begin{center}
	\begin{figure}[!ht]
\includegraphics[scale=0.5]{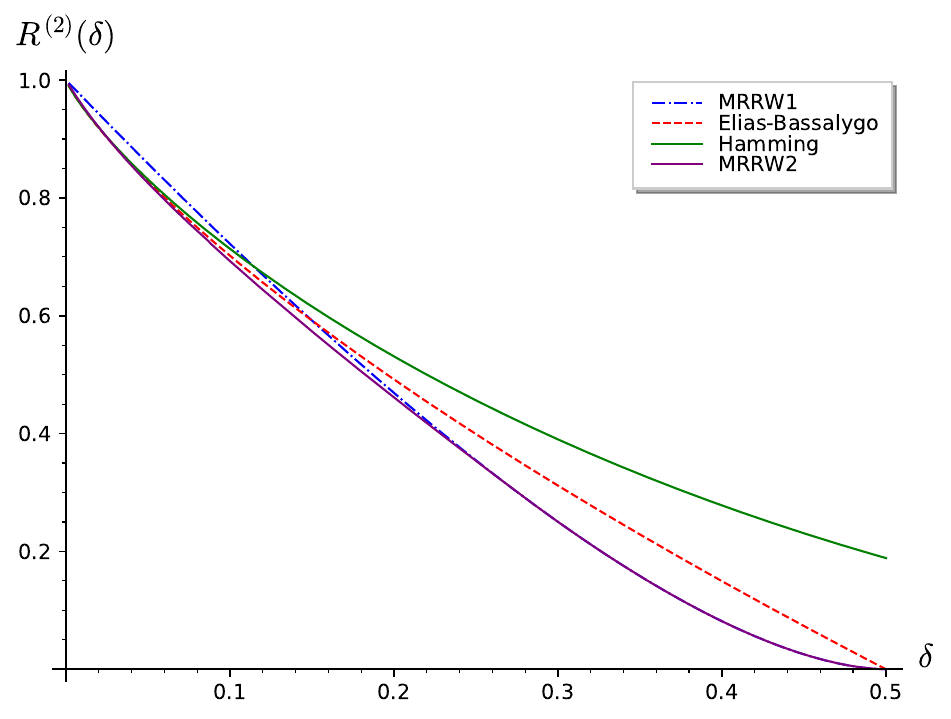}
\caption{Upper bounds over $R^{(2)}(\delta)$ via the linear program with the Hamming (Theorem \ref{theo:H}), Elias-Bassalygo (Theorem \ref{theo:EB}), MRWW1 (Theorem \ref{theo:MRRW1}) bounds and MRRW2 being the second linear programming bound \cite[Eq. (1.4)]{MRRW77}.\label{fig:cube}}
\end{figure}
\end{center}

\begin{center}
	\begin{figure}[!ht]
		\includegraphics[scale=0.5]{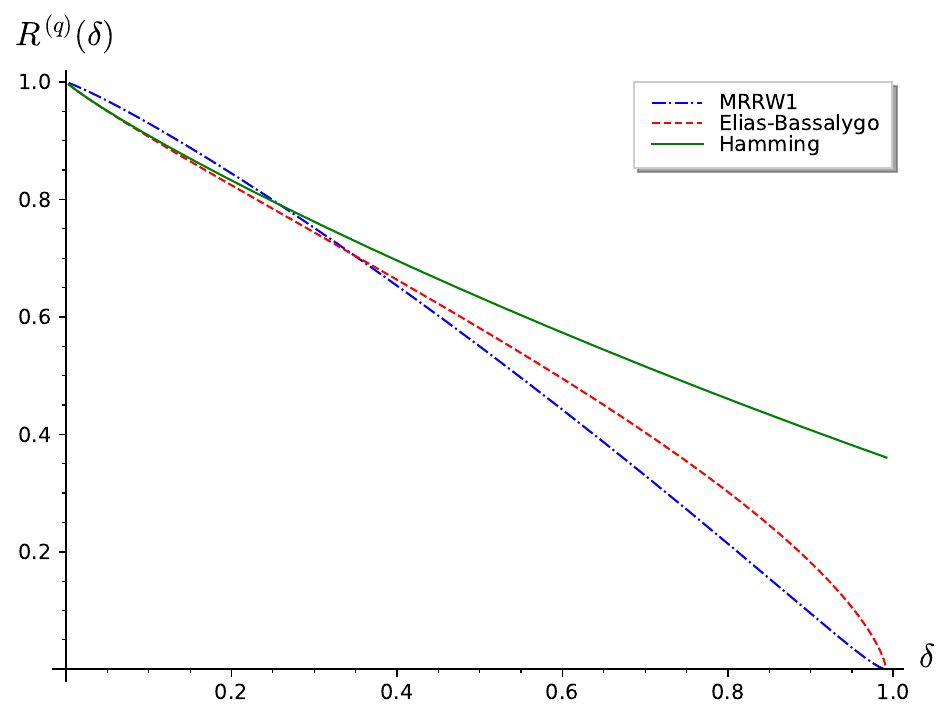}
		\caption{Upper bounds over $R^{(q)}(\delta)$ via the linear program with the Hamming (Proposition \ref{theo:H}), Elias-Bassalygo (Theorem \ref{theo:EB}) and MRWW1 (Theorem \ref{theo:MRRW1}) for $q=121$.\label{fig:hyperCube}}
	\end{figure}
\end{center}

We now recap in Figure \ref{fig:cube} the asymptotic upper bounds over~$R^{(q)}(\delta)$ obtained in Proposition \ref{theo:H} and Theorems \ref{theo:EB},\ref{theo:MRRW1} in the binary case, \ie{} $q=2$. We have also added the best known upper bound on $R(\delta)$: the second linear programming bound from~\cite[Eq. (1.4)]{MRRW77}. \COMMENT{As we mentioned in the introduction, the latter was obtained in \cite{MRRW77} via an upper bound over constant-weight binary codes and not directly via the linear program derived from the Hamming scheme. More precisely, \cite{MRRW77} used first the following bound (which turns out to be the key inequality to obtain the Elias-Bassalygo bound via combinatorial arguments), 
$$
\forall a \in \llbracket 0, \lfloor n/2 \rfloor \rrbracket, \quad A^{(2)}(n,d) \leq \frac{2^{n}}{\binom{n}{a}} \; A(n,d,a). 
$$
Therefore, providing upper bounds on $A^{(2)}(n,d)$ can be reduced to finding upper bounds on $A(n,d,a)$ and then optimizing over the  radius $a$. To obtain good bounds on $A(n,d,a)$, \cite{MRRW77} relied on the linear program derived from the Johnson sphere. We will proceed similarly in the next subsection by instantiating Theorems \ref{theo:Hamming}, \ref{theo:Johnson} and \ref{theo:eigenvalueBound} in this context.}
\COMMENT{Though the second linear programming bound was obtained thanks to a solution of the linear program derived from the Johnson scheme, Rodemich~\cite{Rod80} showed how to turn the latter into a solution of the linear program derived this time from the Hamming scheme. In other words, Rodemich's result shows that there exists a better solution of the linear program in the Hamming scheme than the one obtained in \cite{MRRW77} and ours. However this result only holds in the binary setting. Indeed, in the $q$-ary setting the Johnson scheme does not yield an association scheme, and therefore Delsarte's approach does not apply, \ie{} there are no linear program to solve. But it can be proved that the Elias-Bassalygo bound is better for any $q$ than the bound derived from Theorem \ref{theo:MRRW1} (which corresponds to the first linear programming bound of \cite{MRRW77} instantiated in the $q$-ary case). In other words, as Rodemich's idea does not apply when $q >2$, our work has exhibited for this setting and small minimum distances, a solution which is better than all previously known solutions of the linear program derived from the Hamming scheme. By way of illustration we give in Figure \ref{fig:hyperCube} the asymptotic upper bounds over~$R^{(q)}(\delta)$ obtained in Theorems \ref{theo:H}, \ref{theo:EB} and \ref{theo:MRRW1} for some $q>2$. }

In the next section, we present our results for the binary Johnson scheme.

\subsection{Binary Johnson Scheme}

We consider now the Johnson scheme $(\mathcal{S}_{a}^{n,2},\tau_{\textup{J}} = \tau_{\textup{H}}/2,a)$ where $\mathcal{S}_{a}^{n,2}$ denotes the set of words of Hamming weight $a \in \llbracket 0, \lfloor n/2 \rfloor \rrbracket$ in the Hamming cube $\mathbb{F}_{2}^{n}$. It is a distance induced scheme with a canonical ordering $\Em_0,\dots,\Em_a$ which is also $Q$-polynomial \cite{DL98}. Its valencies and multiplicities are given by,
$$
\forall i \in \llbracket 0,a \rrbracket, \quad v_i = \binom{a}{i}\binom{n-a}{i} \quad , \quad m_i = \binom{n}{i} - \binom{n}{i-1}.
$$

The $q$-numbers of the Johnson schemes involve {\it Hahn polynomials} ${H}_{k}^{n}$ which are defined as, 
$$
\forall k \in \llbracket 0,a \rrbracket, \quad {H}_{k}^{n,a}(X) \eqdef m_k \sum_{j \in \llbracket 0,k \rrbracket} (-1)^{j}\frac{\binom{k}{j}\binom{n+1-k}{j}}{v_j}\binom{X}{j},
$$
where the $v_j$ depend on $a$. More precisely, the $q$-numbers are then the integers given by the evaluation of the Hahn polynomials over~$\llbracket 0,a \rrbracket$, \ie{}
\begin{equation}
	\forall i,k \in \llbracket 0,a \rrbracket, \quad q_{k}(i) \eqdef {H}_{k}^{n,a}(i).
\end{equation}
We have in particular, 
\begin{equation}\label{eq:q1Johnson}
q_1(i) = (n-1)\left(1 - \frac{ni}{a(n-a)}\right)
\end{equation} 
which is a decreasing function as required in Theorems \ref{theo:Johnson} and \ref{theo:eigenvalueBound}. We consider the linear program associated to this association scheme and we denote $A_{\textup{LP}}(n,d,a)$ (as per Subsection ~\ref{Section:LP}) its maximum value. Let,
$$
R_{\textup{LP}}(\delta,\alpha) \eqdef \lim\limits_{n \rightarrow \infty} \frac{1}{n} \log_2  A_{\textup{LP}}(n,\lfloor n\delta \rfloor, \lfloor n \alpha \rfloor)
$$ 
be its asymptotic value. We have,
$$ 
A(n,d,a) \le A_{\textup{LP}}\left(n,\left\lfloor \frac{d}{2} \right\rfloor,a\right) \quad , \quad R(\delta,\alpha) \le R_{\textup{LP}}(\delta/2,\alpha).
$$
The factor two in the distance comes from the fact that in the association scheme $(\mathcal{S}_a^{n,2},\tau_{\textup{J}},a)$, the distance is half of the hamming distance while $A(n,d,a)$ and $R(\delta,\alpha)$ are defined with respect to the Hamming distance. 
\medskip

{\bf \noindent Hamming Bound.} By using the valencies of $(\mathcal{S}_a^{n,2},\tau_{\textup{J}},a)$ and Theorem \ref{theo:Hamming} we easily recover the Hamming bound.

\begin{theorem}[Hamming Bound for $A_{\textup{LP}}(n,d,a)$]\label{theo:JSHB}
Fix integers $n,a,d$, where $a \in \llbracket 0, \lfloor n/2 \rfloor \rrbracket$ and~$d \in \llbracket 1,n \rrbracket$. We have, 
$$
A(n,2d,a) \leq A_{\textup{LP}}(n,d,a) \le \frac{\binom{n}{a}}{\sum_{x=0}^{\left\lfloor \frac{d-1}{2} \right\rfloor}
v_x} = 
	\frac{\binom{n}{a}}{\sum_{x=0}^{\left\lfloor \frac{d-1}{2} \right\rfloor} \binom{a}{x}\binom{n-a}{x}},
$$
which implies asymptotically for any $\alpha \in (0,1/2)$ and $\delta \in (0,\alpha(1-\alpha))$,
$$
R(2\delta,\alpha) \leq R_{\textup{LP}}(\delta,\alpha) \leq  h_2(\alpha) - \left(\alpha h_2\left( \frac{\delta}{2\alpha} \right) + (1-\alpha)h_2\left( \frac{\delta}{2(1-\alpha)} \right)\right).
$$
\end{theorem}

{\bf \noindent Elias-Bassalygo Bound.} We now instantiate to the Hamming sphere $\mathcal{S}^{n,2}_{a}$ our generalized Elias-Bassalygo bound of Theorem \ref{theo:Johnson}. It turns out that this bound is known since the work of Aaltonen~\cite{Aal81}.

\begin{theorem}[Elias-Bassalygo Bound for $A_{\textup{LP}}(n,d,a)$]\label{theo:JSEB}
	Fix integers $n, a, d$, where $a \in \llbracket 0, \lfloor n/2 \rfloor \rrbracket$ and~$d \in \llbracket 0,a(n-a)/n \rrbracket$. We have,
	\begin{multline*} 
	A(n,2d,a) \leq A_{\textup{LP}}(n,d,a) \leq (n-1)\frac{nd}{a(n-a)} \cdot \frac{\binom{n}{a}}{\binom{a}{u}\binom{n-a}{u}}, \\ \textrm{where} \quad u \eqdef  \left\lfloor \frac{a(n-a)}{n} \left( 1 - \sqrt{1-\frac{nd}{a(n-a)}+\frac{1}{n-1}} \right)\right\rfloor
	\end{multline*} 
	which implies asymptotically for any  $\alpha \in (0,1/2)$ and $\delta \in [0,\alpha(1-\alpha)]$,
	$$
	R(2\delta,\alpha) \le R_{\textup{LP}}(\delta,\alpha) \leq 
		h_2(\alpha) - \left( \alpha h_2\left( \frac{K(\delta,\alpha)}{\alpha}\right) + (1-\alpha)h_2\left( \frac{K(\delta,\alpha)}{1-\alpha}\right) \right) 
	$$ 
	where $K(\delta,\alpha) \eqdef \alpha(1-\alpha)\left( 1 - \sqrt{1-\frac{\delta}{\alpha(1-\alpha)}} \right)
	$.
\end{theorem}
\begin{proof}
	Our strategy is to apply Theorem \ref{theo:Johnson}. First $q_{1}$ is indeed a decreasing function. Now, let us compute,
		$$
	u \in \left\{ u_{0}\in \llbracket 0,n \rrbracket: \; \frac{q_{1}(u_{0})^{2}}{q_{1}(0)} \geq q_{1}(d)+1 \right\} \mbox{ which maximizes } \; \binom{a}{u}\binom{u}{n-a}.
	$$
	We have,
	$$
	q_{1}(0)\left( q_{1}(d) + 1 \right) = (n-1)^{2}\left( 1 - \frac{nd}{a(n-a)} + \frac{1}{n-1} \right).
	$$
	Since $d \in \llbracket 0,a(n-a)/n \rrbracket$, the right hand side is nonnegative and 
	$$
	q_{1}(u_{0})^{2} \geq q_{1}(0)\left( q_{1}(d) + 1 \right) \Leftrightarrow 1-\frac{nu_{0}}{a(n-a)} \geq \sqrt{1 - \frac{nd}{a(n-a)} + \frac{1}{n-1}}
	$$
	showing that we have to choose $u$ smaller than,
$$
	\left\lfloor \frac{a(n-a)}{n} \left( 1 - \sqrt{1-\frac{nd}{a(n-a)}+\frac{1}{n-1}} \right)\right\rfloor .
	$$
	We can choose $u$ as above as $y \mapsto \binom{a}{y}\binom{n-a}{y}$ is an increasing function over $\left\llbracket 0,\left\lfloor \frac{a(n-a)}{n} \right\rfloor\right\rrbracket$. 
	Applying Theorem \ref{theo:Johnson} ends the first part of the proof. The asymptotic result easily follows from Lemma~\ref{lemma:binEntropy}. 
\end{proof}

{\bf \noindent MRRW Bound.} We end our instantiations by the bound from Theorem \ref{theo:eigenvalueBound} to the Johnson scheme. We recover the bound from \cite[Eq. (2.16)]{MRRW77}.

\begin{theorem}[MRRW1-type Bound for $A_{\textup{LP}}(n,d,a)$]\label{theo:JSMRRW} 	Fix integers $n, a, d$, where $a \in \llbracket 0, \lfloor n/2 \rfloor \rrbracket$ and~$d \in \llbracket 0,a(n-a)/n \rrbracket$. Let $r$ be an integer such that,
	$$ 
	\zeta_r^{(1)} \le (d-1) < \zeta_{r-1}^{(1)},
	$$
	where $\zeta_x^{(1)}$ is the first zero of $H_r^{n,a}(X)$.
	We have 
	$$
	A(n,2d,a) \leq A_{\textup{LP}}(n,d,a) \le (q_1(0) - q_1(d))\sum_{x = 0}^{r-1} m_r,
	$$
which implies asymptotically  for any 
 $\alpha \in [0,1/2]$ and $\delta \in [0,\alpha(1-\alpha)]$,
	$$
	R(2\delta,\alpha) \le R_{\textup{LP}}(\delta,\alpha) \le h_{2}(B(\delta,\alpha) \quad \textrm{with} \;\;  B(\delta,\alpha) \eqdef \frac{1}{2}\left( 1 - \sqrt{1-4\left(\sqrt{\alpha(1-\alpha)-\delta(1-\delta)}-\delta \right)^{2} }\right). 
	$$
\end{theorem}
\begin{proof}
	Fix integers $n,d,a$.
	We write $q_1(d-1) - q_1(d) = \frac{n(n-1)}{a(n-a)} \ge 1$ which implies,
$$
q_1(d) + 1 \le q_1(d-1).
$$
 For each $x \in \llbracket 0,a \rrbracket$, let $\zeta_x^{(1)} < \zeta_x^{(2)}< \dots < \zeta_{x}^{(a)}$ be the zeros of $H_x^{n,a}(X)$ in $[0,a]$. We know that the zeros of~$H_{x}^{n,a}(X)$ and $H_{x+1}^{n,a}(X)$ are interlaced, \ie{}
 \begin{equation}\label{eq:zetaHahn} 
 \zeta_{x}^{(i-1)} < \zeta_{x+1}^{(i)} < \zeta_{x}^{(i)}.
\end{equation}
 Furthermore, we know that there exists an integer in the open interval $(\zeta_{x+1}^{(i)},\zeta^{(i)}_{x})$ \cite[\S B]{MRRW77}.
Therefore, by supposing that there exists $x$ such that $\zeta_{x}^{(1)} < d-1$, we can choose the minimum $r$ such that,
\begin{align*}
\zeta_{r}^{(1)} \le (d-1) < \zeta_{r-1}^{(1)} < \zeta_{r}^{(2)}
\end{align*}
where in the last inequality we used Equation~\eqref{eq:zetaHahn}.
Let also $r^{\perp} \eqdef d-1$. 
Since, as shown above $r^{\perp} < \zeta_{r}^{(2)}$ we have $H_r^{n,a}(x) \le 0$ for $x \in [\zeta^{(1)}_r,r^{\perp}]$ as $H_{r}^{n,a}(0) = m_{r}>0$ and the zeros of $H_{r}^{n,a}(X)$ are simple. In particular, $q_r(r^{\perp}) = H_r^{n,a}(r^{\perp}) \le 0$. Moreover, we know that $\zeta_{x}^{(1)} > \zeta_{r}^{(1)}$ for $x \in \llbracket 0,r-1 \rrbracket$. Therefore as $H_{x}^{n,a}(0) = m_{x} > 0$, we have that for all $x \in \llbracket 0,r-1 \rrbracket$, $q_{x}(r^{\perp}) = H_{x}^{n,a}(r^{\perp})\ge 0$. One can therefore use Theorem ~\ref{theo:eigenvalueBound} to obtain, 
$$ 
A_{\textup{LP}}(n,d,a) \le (q_1(0) - q_1(d)) \sum_{x = 0}^{r-1} m_{x}.
$$
However, we still need to ensure that there exists $x \in \llbracket 0,a \rrbracket$ such that $\zeta_{x}^{(1)}<d-1$. To this aim we will use the asymptotic of $\zeta_{x}^{(1)}$ for $n$ large enough. First, we choose $d \eqdef \lfloor \delta n \rfloor$ and $a \eqdef \lfloor \alpha n \rfloor$.  Let us suppose that $x/n \mathop{\longrightarrow}\limits_{n \to +\infty} \beta \in [0,\alpha]$. We know from \cite[\S F]{DL98} that,
$$
\frac{\zeta_{x}}{n} \mathop{=}\limits_{n \to +\infty} \zeta(\beta) + o(1),
$$
where,
$$
\zeta(\beta) \eqdef \frac{\alpha(1-\alpha)-\beta(1-\beta)}{1+2\sqrt{\beta(1-\beta)}}.
$$
Furthermore, we know that $\zeta$ maps the interval $[0,\alpha]$ onto $[0,\alpha(1-\alpha)]$. 
We also know that $\zeta$ admits an inverse~$\zeta^{-1}$ that maps $[0,\alpha(1-\alpha)]$ to $[0,\alpha]$ and it is an increasing function. Recall that we supposed $\delta \in [0,\alpha(1-\alpha)]$ where $a = \lfloor \alpha n \rfloor$. Therefore, the minimum $r$ for which we can ensure (for $n$ large enough) $\zeta_{r}^{(1)} < d-1$ is such that $r/n \mathop{\longrightarrow}\limits_{n \to +\infty} \beta$ where, 
$$
\beta = B(\delta,\alpha) \eqdef \zeta^{-1}(\delta) = \frac{1}{2}\left( 1 - \sqrt{1-4\left(\sqrt{\alpha(1-\alpha)-\delta(1-\delta)}-\delta \right)^{2} }\right). 
$$
Therefore, when $\delta \in [0,\alpha(1-\alpha)]$, we obtain as asymptotic bound,
$$
R_{\textup{LP}}(\delta,\alpha) \leq h_{2}\left(B(\delta,\alpha)  \right)
$$
which concludes the proof. 
\end{proof}

\begin{center}
	\begin{figure}[h!]
		\includegraphics[scale=0.5]{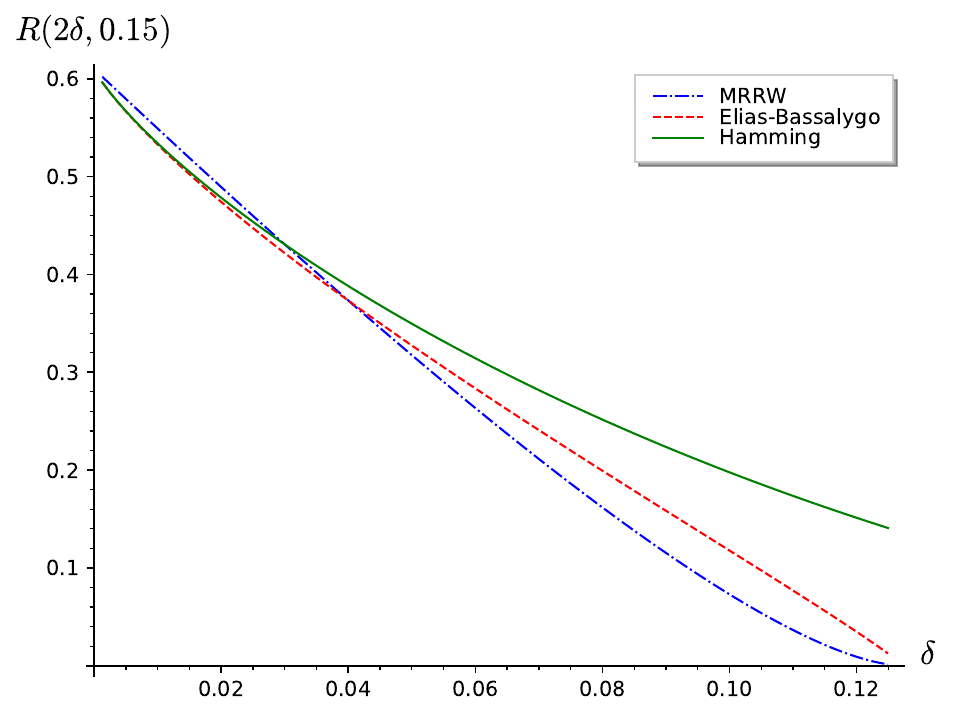}
		\caption{Upper bounds on $R_{LP}(\delta,\alpha)$ via the linear program with the Hamming (Theorem \ref{theo:JSHB}), Elias-Bassalygo (Theorem \ref{theo:JSEB}) and MRRW (Theorem \ref{theo:JSMRRW}) for a relative radius $\alpha = 0.15$.\label{fig:JS}}
	\end{figure}
\end{center}

{\bf \noindent Discussion.} We depict in Figure \ref{fig:JS} the asymptotic upper bounds over $R(\delta,\alpha)$ obtained in Theorems \ref{theo:JSHB}, \ref{theo:JSEB} and \ref{theo:JSMRRW} for some relative radius $\alpha$. As it can be noticed the generalized Elias-Bassalygo bounds gives better result than the MRRW-like bound from Theorem \ref{theo:JSMRRW}. It turns out that this result holds for $\delta$ close to $0$ for any relative radius $\alpha$. Indeed, we can compute \begin{align*}
	B(\delta,\alpha) \mathop{=}\limits_{\delta \rightarrow 0^+} \alpha - \frac{1 + 2\sqrt{\alpha - \alpha^2}}{1 - 2\alpha}\delta + o(\delta) 
\end{align*}
which gives,
$$
R_{\textup{MRRW}}(\delta,\alpha) \eqdef h_2(B(\delta,\alpha)) \mathop{=}\limits_{\delta \rightarrow 0^+} h_2(\alpha) - \frac{(\log_2(1-\alpha) - \log_2 \alpha)(1 + 2\sqrt{\alpha - \alpha^2})}{1 - 2\alpha}\delta + o(\delta).
$$
On the other hand, we have $K(\delta,\alpha) \mathop{=}\limits_{\delta \rightarrow 0^+} \frac{\delta}{2} + o(\delta)$ and $h_2(\delta) \mathop{=}\limits_{\delta \rightarrow 0^+} -\delta\log_2 \delta - o(\delta\log_2 \delta)$ which gives 
$$ 
R_{\textup{EB}}(\delta,\alpha) \eqdef h_2(\alpha) - \alpha h_2\left( \frac{K(\delta,\alpha)}{\alpha}\right) - (1-\alpha)h_2\left( \frac{K(\delta,\alpha)}{1-\alpha}\right) \mathop{=}\limits_{\delta \rightarrow 0^+} h_2(\alpha) + \delta\log_2(\delta) + o(\delta\log_2(\delta)).
$$
One can see here, that $R_{\textup{EB}}(\delta,\alpha)$ decreases faster for $\delta \longrightarrow 0^+$ than for the MRRW bound~$R_{\textup{MRRW}}(\delta,\alpha)$, and this holds for any $\alpha \in (0,\frac{1}{2})$. 

	\section*{Acknowledgments.}
	The authors are particularly grateful to Alexander Barg for his insightful remarks on this article, in particular for bringing many references to their attention. The work of Thomas Debris-Alazard was funded by the French Agence Nationale de la Recherche through ANR JCJC COLA (ANR-21-CE39-0011). The authors were supported by  the PEPR quantique France 2030 programme (ANR-22-PETQ-0008).

	\bibliographystyle{alpha}
	\addcontentsline{toc}{section}{Bibliography}
	\newpage
		\appendix
	\section{Proofs on Claimed Results about Association Scheme}\label{app:proofs}

Our aim in this appendix is to provide a self-contain section proving all results from Subsections~\ref{subsec:equiAssoc} to \ref{subsec:pointWiseE}. We recall here definitions and propositions of these subsections instead of referring to them for ease of reading. Furthermore, we will use Dirac's {\it bra-ket} notation as introduced in Subsection~\ref{subsec:code:DualWD}. 
\medskip

 Recall that where are given $(\X,\tau,n)$, where $\X$ is a finite set and $\tau : \X^{2} \longrightarrow \llbracket 0,n \rrbracket$ is a distance. Furthermore, we consider the following adjacency matrices $\vec{D}_{i} \in \mathbb{C}(\X^{2})$ for $i \in \llbracket 0,n \rrbracket$, 
$$
\vec{D}_{i}(x,y) \eqdef \begin{cases}
	1 \mbox{ if } \tau(x,y) = 1 \\
	0 \mbox{ otherwise}
\end{cases}. 
$$    
In other words,
\begin{equation}\label{eq:Diketbra}
	\vec{D}_{i} = \sum_{\substack{x,y \in \X \\ \tau(x,y) = i}} \ketbra{x}{y} .
\end{equation} 
We have defined distance induced association schemes as triplets $(\X,\tau,n)$ satisfying the equipartition and non-degenerate properties which are defined as follows.

\eqprop*

From the symmetry of $\tau$ we easily get the following equation,
\begin{equation}\label{eq:pijksym}
	\forall i,j,k \in \llbracket 0,n \rrbracket, \quad p_{i,j}^{k} = p_{j,i}^{k}. 
\end{equation}

The equipartition property ensures that the complex vector space generated by the adjacency matrices $\vec{D}_{i}$ is closed under matrix multiplication, \ie{} it forms an associative algebra. 

\propoMultd*

\begin{proof}
	According to Equation \eqref{eq:Diketbra} we have the following computation,
	\begin{align*}
		\vec{D}_{i}\cdot \vec{D}_{j}&= \sum_{\substack{x,y,x',y' \in \X \\ \tau(x,y) = i \mbox{ \scriptsize and } \tau(x',y') = j}} \ketbra{x}{y} \cdot \ketbra{x'}{y'}\\
		&= \sum_{\substack{x,z,y' \in \X \\ \tau(x,z) = i \mbox{ \scriptsize and } \tau(z,y') = j}} \ketbra{x}{y'} \\
		&= \sum_{k\in \llbracket 0,n \rrbracket} \sum_{\substack{x,y' \in \X \\ \tau(x,y')=k }} \; \sum_{\substack{z \in \X \\ \tau(x,z) = i \mbox{ \scriptsize and } \tau(z,y') = j}} \ketbra{x}{y'} \\
		&=  \sum_{k \in \llbracket 0,n \rrbracket}\sum_{\substack{x,y' \in \X \\ \tau(x,y')=k }} p_{i,j}^{k} \ketbra{x}{y'}
	\end{align*}
	where in the last equality we used the definition of the $p_{i,j}^{k}$ given in Definition \ref{def:pijk}. To conclude the proof it remains to use Equation~\eqref{eq:Diketbra}. 
\end{proof}

We are now ready to properly define distance induced association schemes (usually known as metric schemes).

\assocScheme*

\subsection{Polynomial Relations}

According to Proposition \ref{propo:multD} and Equation \eqref{eq:pijksym}, the associated matrices $\vec{D}_{i}$'s from a given association scheme $(\X,\tau,n)$ commute. Therefore these adjacency matrices (over $\mathbb{C}$) are diagonalizable in the same basis (they are all diagonalizable as $\vec{D}_{i}^{\dagger} = \vec{D}_{i}$ by symmetry of the distance $\tau$). We can actually prove that they also share the same eigenspaces.

Notice that $\tau$ is ranging over $\llbracket 0,n \rrbracket$ and it satisfies the triangular inequality (it is a distance) from which we deduce the crucial equation,
\begin{equation}\label{eq:recDi} 
	p_{i,j}^{k} = 0 \; \mbox{ if }\;  k > i+j,  \mbox{ or } \; |j-i| > k \; \mbox{ or as soon as} \; i,j,k > n .
\end{equation}
Notice that together with Proposition \ref{propo:multD} it implies the fundamental relation,
\begin{equation}\label{eq:order3Di} 
	\forall k \in \llbracket 0,n \rrbracket, \quad \vec{D}_{1}\vec{D}_{k} = p_{1,k}^{k-1} \vec{D}_{k-1} + p_{1,k}^{k}\vec{D}_{k} + p_{1,k}^{k+1}\vec{D}_{k+1} .
\end{equation} 
But, as $(\X,\tau,n)$ is non-degenerate, \ie{} $p_{1,k}^{k+1} \neq 0$, 
\begin{equation}\label{eq:Dkplus1} 
	\vec{D}_{k+1} = \frac{1}{p_{1,k}^{k+1}}\left( \vec{D}_{1}\vec{D}_{k} - p_{1,k}^{k-1} \vec{D}_{k-1} - p_{1,k}^{k}\vec{D}_{k} \right).
\end{equation} 
In particular, $\vec{D}_{2}$ is some polynomial of degree $1$ in $\vec{D}_{1}$ ($\vec{D}_{0}$ is the identity matrix).
We can then extend this result to the other $\vec{D}_{i}$'s as shown in the following proposition.

\begin{proposition}\label{propo:Dipoly}
	Let $(\X,\tau,n)$ be a distance induced association scheme with adjacency matrices~$(\vec{D}_{i})_{i \in \llbracket 0,n\rrbracket}$. Then, for all $i \in \llbracket 0,n \rrbracket$, there exists a polynomial $P_{i} \in \mathbb{R}[X]$ of degree $i$ with leading coefficient $\left( \prod_{j=1}^{i-1} p_{1,j}^{j+1} \right)^{-1}$ such that 
	$$
	\vec{D}_{i} = P_{i}\left( \vec{D}_{1} \right) 
	$$
	We call these polynomials the fundamental $P$-polynomials of the association scheme $(\X, \tau,n)$.
\end{proposition}

\begin{proof}
	This proposition follows from a straightforward induction using Equation \eqref{eq:Dkplus1} and the fact that $(\X,\tau,n)$ is non-degenerate.
\end{proof}

\subsection{Eigenstates and Eigenvalues of the $\vec{D}_i$'s: Introducing the $\vec{E}_i$'s}

The polynomial relation from Proposition \ref{propo:Dipoly} shows that the $\vec{D}_{i}$'s share common eigenspaces in addition to be co-diagonalizable. Furthermore, combining this result with Proposition \ref{propo:multD} (in particular Equation~\eqref{eq:recDi}) shows that the $\vec{D}_{i}$'s can be decomposed as the sum of $n+1$ orthogonal projectors.

\begin{proposition}\label{propo:Divp}
	Let $(\X,\tau,n)$ be a distance induced association scheme with associated adjacency matrices~$(\vec{D}_{i})_{i \in \llbracket 0,n \rrbracket}$. There exist orthogonal projectors $(\vec{E}_{i})_{i \in \llbracket 0,n \rrbracket}$ and distinct $(\lambda_{i})_{i\in \llbracket 0,n \rrbracket} \in \mathbb{R}^{n+1}$
such that,
$$
	\forall i \in \llbracket 0,n \rrbracket, \quad \vec{D}_{i} = \sum_{j \in \llbracket 0,n \rrbracket} P_{i}(\lambda_{j})\vec{E}_{j}
	$$ 
	where the $P_{i}$'s are the fundamental $P$-polynomials of the association scheme $(\X, \tau,n)$.

	Furthermore, one can choose
	$$
	\vec{E}_{0} = \frac{1}{{|\X|}} \sum_{x,y \in \llbracket 0,n \rrbracket} \ketbra{x}{y}.
	$$
\end{proposition}
\begin{proof}
	First, $\vec{D}_{1}$ is diagonalizable because $\vec{D}_{1}^{\dagger} = \vec{D}_{1}$ and its eigenvalues are real. Therefore there exist real numbers $\lambda_{0} > \lambda_{1} > \cdots > \lambda_{m}$ and orthogonal projectors $\vec{E}_{0},\dots, \vec{E}_{m}$ such that,
	\begin{equation}\label{eq:DiPiEi}
		\vec{D}_{1} = \sum_{j\in \llbracket 0,m \rrbracket} \lambda_{j}\vec{E}_{j}.
	\end{equation}
	From Proposition \ref{propo:Dipoly} we deduce that (notice that $\vec{E}_{j}^{2} = \vec{E}_{j}$ and $\vec{E}_{i}\vec{E}_{j} = \vec{0}$ for $i \neq j$),
	\begin{equation}\label{eq:DiPiD1proof}
		\forall i \in \llbracket 0,n \rrbracket, \quad \vec{D}_{i} = P_{i}(\vec{D}_{1}) = \sum_{j \in \llbracket 0,m \rrbracket} P_{i}(\lambda_{j}) \vec{E}_{j}
	\end{equation} 
	where the $P_{i}$'s are the fundamental $P$-polynomials. They have degree $i$ and leading coefficient~$\left( \prod_{j=1}^{i-1} p_{1,j}^{j+1} \right)^{-1}$. Let us show that $m = n$ to conclude the proof.

	First, $m \geq n$. Indeed, by the decomposition from Proposition \ref{propo:multD} and using Equation \eqref{eq:recDi} we get,
	\begin{equation}\label{eq:D1Dn} 
	\vec{D}_{1}\vec{D}_{n} = p_{1,n}^{n-1}\vec{D}_{n-1} + p_{1,n}^{n}\vec{D}_{n}.
\end{equation}
	Let, 
	$$
	Q \eqdef P_1P_n - p_{1,n}^{n-1}P_{n-1} - p_{1,n}^{n}P_{n} \in \mathbb{R}[X].
	$$
	This polynomial has degree $\leq n+1$. Furthermore, plugging $\vec{D}_{n} = P_{n}(\vec{D}_1)$ and $\vec{D}_{1}= P_{1}(\vec{D}_{1})$ into Equation \eqref{eq:D1Dn} shows that $Q(\vec{D}_{1}) = 0$. Therefore, from the fact that the $\vec{E}_{i}$'s are orthogonal projectors and Equation \eqref{eq:DiPiEi},
	$$
	\forall i \in \llbracket 0,m \rrbracket, \quad Q(\lambda_i) = 0 .
	$$ 
	But the degree of $Q$ is $\leq n+1$ showing that $m \leq n$. 
	Let us show now that $n \leq m$. Assume by contradiction that $m<n$. Let $R(X) \eqdef \prod_{i=0}^{m}(X-\lambda_i)$. By Equation \eqref{eq:DiPiEi}, 
	$$
	R(\vec{D}_1) = \vec{0} .
	$$ 
	By writing $R(X) = X^{m} + \sum_{j=0}^{m-1} a_j X^{j}$, we obtain 
	\begin{equation*}
		R(\vec{D}_1) = \vec{D}_{1}^{m} + \sum_{j=0}^{m-1} a_{j}\vec{D}_{1}^{j} .
	\end{equation*}
	By definition, $P_{m}$ has leading coefficient $\left( \prod_{j=1}^{m-1} p_{1,j}^{j+1} \right)^{-1}$. Therefore, by using $\vec{D}_{m} = P_{m}(\vec{D}_{1})$ (see Equation \eqref{eq:DiPiD1proof}) with the fact that the $\vec{E}_{j}$'s are orthogonal projectors, we obtain for some $b_{j}$'s,
	$$
	R(\vec{D}_{1})= \left( \prod_{j=1}^{m-1} p_{1,j}^{j+1}\right) \vec{D}_{m} + \sum_{j=0}^{m-1} b_j \vec{D}_{j} .
	$$
It shows $R(\vec{D}_1) \neq \vec{0}$ which is a contradiction. Indeed, $(\X,\tau,n)$ is non-degenerate: by definition the~$p_{1,j}^{j+1}$'s are non-zero.

To conclude the proof let us show (up to a re-ordering) that we have,
$$
\vec{E}_{0} = \frac{1}{|\X|} \sum_{x,y \in X} \ketbra{x}{y} = \frac{1}{|\X|} \vec{J}. 
$$
First, notice that $\vec{J} = \sum_{i \in \llbracket 0,n \rrbracket} \vec{D}_{i}$. Therefore, $\vec{J}$ belongs to the space generated by the $\vec{D}_{i}$'s which is also generated by the $\vec{E}_{i}$'s. It shows that we can write $\vec{J} = \sum_{i \in \llbracket 0,n \rrbracket} \beta_{i} \vec{E}_{i}$ and we can suppose that $\beta_{0}\neq 0$. As the $\vec{E}_{i}$'s are orthogonal projectors we get,
\begin{equation}\label{eq:E0J} 
\vec{E}_{0} = \frac{1}{\beta_{0}}\vec{J} \vec{E}_{0} = \vec{E}_{0}\vec{J}.
\end{equation}
In particular $\vec{E}_{0}$ and $\vec{J}$ commute.
On the other hand, as $\vec{J}^{2} =|\X| \; \vec{J}$, we also have,
$$
\vec{J} \vec{E}_{0} = \frac{1}{|\X|} \vec{J}^{2} \vec{E}_{0} = \frac{1}{|\X|} \vec{J} \left( \vec{E}_{0} \vec{J} \right)= \frac{\beta}{|\X|} \vec{J} 
$$
where $\beta$ is the sum of all the entries of $\vec{E}_{0}$. We deduce by combining the above equation and Equation \eqref{eq:E0J} that $\vec{E}_{0}$ is a scalar multiple of $\vec{J}$. Using now that~$\vec{E}_{0}^{2} = \vec{E}_{0}$ shows that $\vec{E}_{0} = 1/|\X| \vec{J}$ which concludes the proof. 
\end{proof}

The fundamental parameters of an association scheme are defined with respect to an ordering of the matrices $(\Em_i)_{i \in \llbracket 0,n \rrbracket}$. In what follows we will only enforce an ordering such that,~$\vec{E}_{0} = {1}/{{|\X|}} \sum_{x,y \in \llbracket 0,n \rrbracket} \ketbra{x}{y}$ which is possible as shown in the above proposition. 

\pnumbers*

\begin{remark} The matrices $\vec{E}_{i}$'s are the orthogonal projectors over the (common) eigenspaces of the adjacency matrices $\vec{D}_{i}$'s. Therefore they sum to the identity, \ie{} $\sum_{i \in \llbracket 0,n \rrbracket} \vec{E}_{i} = \vec{Id}$. But $\vec{D}_{0} = \vec{Id}$ which shows from the decomposition in Definition \ref{def:pnumber} that,
\begin{equation}\label{eq:p0}
	\forall j \in \llbracket 0,n \rrbracket, \quad p_{0}(j) = 1 .
\end{equation}
\end{remark}

The $p$-numbers are well defined as matrices $(\vec{D}_{i})_{i \in \llbracket 0,n \rrbracket}$ and $(\vec{E}_{i})_{i \in \llbracket 0,n \rrbracket}$ generate the same Hilbert space of dimension $n+1$.

\begin{proposition}\label{propo:p3order}
We have,
$$
p_{1}(j) p_{i}(j) = p_{1,i}^{i-1}p_{i-1}(j) + p_{1,i}^{i}p_{i}(j) + p_{1,i}^{i+1}p_{i+1}(j) 
$$
\end{proposition}
\begin{proof}
First, by using Definition \ref{def:pnumber},
$$
\vec{D}_{1}\cdot \vec{D}_{i} = \sum_{j,\ell \in \llbracket 0,n \rrbracket} p_{1}(j)p_{i}(\ell)\vec{E}_{i}\cdot \vec{E}_{\ell} = \sum_{j \in \llbracket 0,n \rrbracket} p_{1}(j)p_{i}(j) \vec{E}_{j}
$$
where we used in the last equality that $\vec{E}_{i}\cdot \vec{E}_{j} = \delta_{i}^{j} \cdot \vec{E}_{i}$ as orthogonal projectors. Recall now that from Equation \eqref{eq:order3Di},
\begin{align*} 
	\forall j \in \llbracket 0,n \rrbracket, \quad \vec{D}_{1}\vec{D}_{i} &= p_{1,i}^{i-1} \vec{D}_{i-1} + p_{1,i}^{i}\vec{D}_{i} + p_{1,i}^{i+1}\vec{D}_{i+1} \\
	&= \sum_{j \in \llbracket 0,n \rrbracket} \left( p_{1,i}^{i-1}p_{i-1}(j) + p_{1,i}^{i}p_{i}(j) + p_{1,i}^{i+1}p_{i+1}(j) \right) \vec{E}_{j}.
\end{align*} 
It ends the proof by using the unicity of the decomposition given in the basis~$(\vec{E}_{i})_{i \in \llbracket 0,n\rrbracket}$. 
\end{proof}

The fact that the $(\vec{D}_{i})_{i \in \llbracket 0,n \rrbracket}$ and $(\vec{E}_{i})_{i \in \llbracket 0,n \rrbracket}$ generate the same Hilbert  enables to define the $q$-numbers, an equivalent of the $p$-numbers (according to Definition \ref{def:pnumber}), where the $\vec{E}_{i}$'s and $\vec{D}_{i}$'s are interchanged.

\qnumbers*

Notice that the $q$-numbers are uniquely defined as for $p$-numbers and they are real numbers as~$\vec{E}_{i}^{\dagger} = \vec{E}_{i}$ and $\vec{D}_{i}^{\dagger} = \vec{D}_{i}$.

\begin{remark}
In Proposition \ref{propo:Divp} we chose an ordering such that 
\begin{equation}\label{eq:E0} 
\vec{E}_{0} = \frac{1}{{|\X|}} \sum_{x,y \in \llbracket 0,n \rrbracket} \ketbra{x}{y} .
\end{equation} 
Here the $\vec{D}_{j}$'s are adjacency matrices associated to a metric. In particular they sum to $|\X| \cdot \vec{E}_{0}$. Therefore it is necessary that, 
\begin{equation*}\label{eq:qq0}
	\forall j \in \llbracket 0,n \rrbracket, \quad q_{0}(j) = 1.
\end{equation*} 
\end{remark}

It may be tempting to conjecture that the $q$-numbers verify also a 3-term order recurrence as the one given for the $p$-numbers in Proposition \ref{propo:p3order}. It will turn out that such relation is crucial for our purpose. However, we first need to define an equivalent of the $p_{i,j}^{\ell}$'s: the $q_{i,j}^{\ell}$'s. There will be defined in Subsection \ref{subsec:pointWiseEE} and they are known as {\it Krein parameters}. To this aim let us study the orthogonality relations of the $\vec{D}_{i}$'s and $\vec{E}_{i}$'s.

	\subsection{Orthogonality Relations}

	It turns out (by symmetry of the underlying distance $\tau$) that the $\vec{D}_{i}$'s are orthogonal, \ie{} $\braket{\vec{D}_i}{\vec{D}_j}= 0$, but also the $\vec{E}_{i}$'s, \ie{} $\braket{\vec{E}_{i}}{\vec{E}_{j}} = 0$, as orthogonal projectors. 	
	
	Recall that we have defined their norms (with a normalization) as follows.

	\normvmi*

	The $p$ and $q$-numbers were derived from matrices $\vec{D}_{i}$'s and $\vec{E}_{i}$'s. They satisfy the following ``orthogonality'' relations.

	\begin{proposition}\label{propo:orthopq}
		For any $i,j \in \llbracket 0,n \rrbracket$, 
		$$
		\sum_{k\in \llbracket 0,n \rrbracket} p_{i}(k)  {p_{j}(k)} m_k = \delta_{i}^{j} \cdot v_{i} \cdot |\X| \quad \mbox{and} \quad  \sum_{k \in \llbracket 0,n \rrbracket} q_{i}(k) {q_{j}(k)} v_k = \delta_{i}^{j} \cdot m_{i} \cdot |\X|
		$$
	\end{proposition} 
	\begin{proof}
		We just have to use Definitions \ref{def:pnumber}, \ref{def:qnumber} and the orthogonality of the~$\vec{D}_{i}$'s and $\vec{E}_{j}$'s.
	\end{proof}

	Furthermore, $p$ and $q$-numbers are related as follows.

	\proposcalProductEiDi*
	
	\begin{proof}
		First, by Definition \ref{def:pnumber} and orthogonality of the $\vec{E}_{j}$'s,
		\begin{equation}\label{eq:scalProdEiDi}
		\forall j \in \llbracket 0,n \rrbracket, \quad \braket{\vec{D}_{i}}{\vec{E}_{j}} = p_{i}(j)\| \vec{E}_{j} \|^{2} = p_{i}(j)m_j \quad \mbox{and} \quad  \braket{\vec{E}_{j}}{\vec{D}_{i}} = \frac{1}{|\X|}q_{j}(i)\| \vec{D}_{i} \|^{2} = q_{j}(i)v_i. 
		\end{equation} 
		To end the proof we just have to use the fact that $\braket{\vec{A}}{\vec{B}}= \overline{\braket{\vec{B}}{\vec{A}}}$ and that the $p_{i}(j)$'s and $q_{i}(j)$'s are real numbers. 
	\end{proof}

		\subsection{Algebra Structure for Pointwise Multiplication}\label{subsec:pointWiseEE}

	In the above subsections we investigated the structure of distance induced association schemes via the complex complex commutative algebra $\mathcal{H}$ generated by its underlying adjacency matrices $\vec{D}_{i}$'s. As we have shown, it turns out that $\mathcal{H}$ is also generated by matrices  $\vec{E}_{i}$'s which are orthogonal as the $\vec{D}_{i}$'s. However, though~$\mathcal{H}$ forms a an algebra for the standard matrix-product, it is also (surprisingly) closed under the pointwise multiplication $(\vec{M},\vec{N}) \mapsto \vec{M} \circ \vec{N}$ where,
	$$
	\vec{M} \circ \vec{N}(x,y) = \vec{M}(x,y)\vec{N}(x,y) .
	$$  
	Indeed the $\vec{D}_{i}$'s verify the following relation 
\begin{equation}\label{eq:Diproj}  
		\vec{D}_{i} \circ \vec{D}_{j} = \delta_{i}^{j} \cdot \vec{D}_{i} 
	\end{equation} 
	and we have the following proposition which in particular gives an equivalent of the $p_{i,j}^{\ell}$'s.

	\begin{proposition}\label{propo:qijk}
We have,  
		$$
		|\X| \cdot \vec{E}_{i}\circ \vec{E}_{j} = \sum_{k\in\llbracket 0,n \rrbracket} q_{i,j}^{k} \vec{E}_{k}, \quad \mbox{ where } q_{i,j}^{k} \eqdef \frac{1}{|\X|} \sum_{m \in \llbracket 0,n \rrbracket} q_{i}(m)q_{j}(m)p_{m}(k) .
		$$
	\end{proposition}
	
	\begin{proof}
First, 
		\begin{align*} 
			\vec{E}_{i}\circ \vec{E}_{j} &= \sum_{x,y\in \llbracket 0,n \rrbracket} \bra{x}\vec{E}_{i}\ket{y} \; \bra{x}\vec{E}_{j}\ket{y}\; \ketbra{x}{y} \\
			&= \frac{1}{|\X|^{2}} \sum_{x,y\in \llbracket 0,n \rrbracket} \left( \sum_{k\in \llbracket 0,n \rrbracket} \bra{x}  q_{i}(k)\vec{D}_{k}\ket{y} \right)  \left( \sum_{\ell \in \llbracket 0,n \rrbracket} \bra{x} q_{j}(\ell)\vec{D}_{\ell}\ket{y} \right) \ketbra{x}{y} \quad \mbox{(By Definition \ref{def:qnumber})} \\
			&=\frac{1}{|\X|^{2}} \sum_{m \in \llbracket 0,n \rrbracket }\sum_{\substack{x,y\in \llbracket 0,n \rrbracket\\\tau(x,y)=m}} q_{i}(m)q_{j}(m)\ketbra{x}{y}
		\end{align*} 
		
		Therefore,
		$$
		\vec{E}_{i}\circ \vec{E}_{j} = \frac{1}{|\X|^{2}}\sum_{m\in \llbracket 0,n \rrbracket} q_{i}(m)q_{j}(m) \vec{D}_{m} = \frac{1}{|\X|^{2}} \sum_{m,\ell \in \llbracket 0,n \rrbracket} q_{i}(m)q_{j}(m)p_{m}(\ell)\vec{E}_{\ell}
		$$
		which concludes the proof. 
\end{proof}

	The numbers $q_{i,j}^{k}$'s are analogous to the $p_{i,j}^{k}$'s but when decomposing in the basis given by the $\vec{E}_{i}$'s and considering the pointwise multiplication. It turns out that the $q_{i,j}^{k}$'s are known as the {\it Krein parameters} of the underlying association scheme and they indeed share the same kind of property. It is readily verified that from Proposition \ref{propo:qijk} that the $q_{i,j}^{\ell}$'s are symmetric, \ie{}
	$$
	\forall i,j,k \in \llbracket 0,n \rrbracket, \quad q_{i,j}^{k} = q_{j,i}^{k} .
	$$
	Furthermore, Krein parameters verify numerous relations. In the following proposition we give some of them which are useful four our purpose. 
	\begin{proposition} Let $(\X,\tau,n)$ be a distance induced association scheme with an ordering~$\vec{E}_{0},\dots, \vec{E}_{n}$. We have, for all $x,y \in \llbracket 0,n \rrbracket$, 
	$$
	(1) \;\; q_{x,x}^{0} = m_{x} > 0\quad , \quad (2)\;\; \sum_{y \in \llbracket 0,n \rrbracket} q_{y,1}^{x} = q_{1}(0)\quad , \quad (3)\;\; m_{x} \cdot q_{y,1}^{x} = m_{y} \cdot q_{x,1}^{y}. 
	$$
	\end{proposition}
	\begin{proof}
		Let us first prove $(1)$. By Proposition \ref{propo:qijk},
		\begin{align*}
			q_{x,x}^{0} &= \frac{1}{|\X|} \sum_{m \in \llbracket 0,n \rrbracket} q_{x}(m)q_{x}(m)p_{m}(0) \\
			&= \frac{1}{|\X|} \sum_{m \in \llbracket 0,n \rrbracket} q_{x}(m)q_{x}(m) \frac{v_{m}}{m_{0}} q_{0}(m) \quad \left(\mbox{By Proposition \ref{propo:scalProductEiDi}}\right)\\
			&= \frac{1}{|\X|}  \sum_{m \in \llbracket 0,n \rrbracket} q_{x}(m)q_{x}(m) v_{m} \quad \left(\mbox{$m_{0} = \rank(\vec{E}_{0}) = 1$ by Equation \eqref{eq:E0}}\right) \\
			&= m_{x}
		\end{align*}
		where in the last equality we used Proposition \ref{propo:orthopq}. We prove now $(2)$. First, according to Proposition \ref{propo:qijk},
		\begin{align*} 
			\sum_{ y \in \llbracket 0,n \rrbracket} q_{y,1}^{x} &= \sum_{y,m \in \llbracket 0,n \rrbracket} q_{1}(m)q_{y}(m)p_{m}(x) \\
			&= \sum_{m \in \llbracket 0,n \rrbracket} q_{1}(m)p_{m}(x)\; \frac{1}{v_{m}}\; \left( \sum_{y \in \llbracket 0,n \rrbracket}  m_{y} p_{m}(y) \right) \\
			&= \sum_{m \in \llbracket 0,n \rrbracket} q_{1}(m)p_{m}(x)\; \frac{1}{v_{m}}\; \left( \sum_{y \in \llbracket 0,n \rrbracket}  m_{y} p_{m}(y)p_{0}(y) \right) \qquad \left(\mbox{By Equation \eqref{eq:p0}}\right) \\
			&= \sum_{m \in \llbracket 0,n \rrbracket}q_{1}(m)p_{m}(x)\; \frac{\delta_{0}^{m}v_{m}}{v_{m}} \qquad \left(\mbox{By Proposition \ref{propo:orthopq}}\right) \\
			&= q_{1}(0)
		\end{align*} 
		where in the last equality we used that $p_{0}$ is constant and equal to $1$ as shown in Equation \eqref{eq:p0}.

		Let us now finish the proof by proving $(3)$. According once again to Proposition \ref{propo:qijk},
		$$
			q_{x,1}^{y} = \sum_{m \in \llbracket 0,n \rrbracket} q_{1}(m)q_{x}(m)p_{m}(y).
		$$
		Therefore, according to Proposition \ref{propo:scalProductEiDi},
			$$
		q_{x,1}^{y} = \sum_{m \in [0,n]} q_{1}(m) \; p_{m}(x) \frac{m_{x}}{v_{m}} \; q_{y}(m) \frac{v_{m}}{m_{y}} 
		= \frac{m_{x}}{m_{y}}\; q_{y,1}^{x} 
		$$
		which ends the proof. 
	\end{proof}
	More surprisingly, Krein parameters are also positive (which is crucial for our purpose, in particular to prove Proposition \ref{propo:positivityOstar}) as shown in the following proposition.

	\positivityKrein*

	\begin{proof}
		Let $\ket{\psi} \in \mathbb{C}(\X)$. Let us introduce the following linear operator,
$$
		\Delta \eqdef \sum_{x \in \X} \braket{\psi}{x} \ketbra{x}{x}
		$$
		We have the following computation, 
		\begin{align}
			\| \vec{E}_{i}\Delta\vec{E}_{j} \|^{2} =  \tr\left( \Delta^{\dagger} \vec{E}_{i}\Delta\vec{E}_{j} \right)
			&= \sum_{ y \in \X} \bra{y} \Delta^{\dagger} \vec{E}_{i}\Delta\vec{E}_{j} \ket{y} \nonumber \\
			&= \sum_{y \in \X} \overline{\braket{\psi}{y}} \bra{y} \vec{E}_{i} \left( \sum_{x \in \X} \braket{\psi}{x}\ketbra{x}{x} \right)\vec{E}_{j}\ket{y} \nonumber \\
			&= \sum_{y,x \in \X}  \overline{\braket{\psi}{y}}\braket{\psi}{x} \vec{E}_{i}(y,x) \vec{E}_{j}(x,y)\nonumber \\
			&= \sum_{y,x \in \X}  \overline{\braket{\psi}{y}}\braket{\psi}{x} \vec{E}_{i}(x,y) \vec{E}_{j}(x,y) \quad \left(\mbox{$\vec{E}_{i}^{\dagger} = \vec{E}_{i}$ and $\vec{E}_{i}$ is real}\right) \nonumber \\
			&= \sum_{k \in \llbracket 0,n \rrbracket} \sum_{y,x \in \X}  q_{i,j}^{k} \overline{\braket{\psi}{y}}\braket{\psi}{x} \vec{E}_{k}(x,y) \label{eq:EiDeltaEj}
		\end{align}
		Furthermore, we have,
		\begin{align*}
			\| \vec{E}_{k}\ket{\psi} \|^{2} = \bra{\psi}\vec{E}_{k}\ket{\psi} 
			&= \bra{\psi} \left( \sum_{x,y \in \X} \vec{E}_{k}(x,y) \ketbra{x}{y} \right) \ket{\psi} = \sum_{y,x \in \X}  q_{i,j}^{k} \overline{\braket{\psi}{y}}\braket{\psi}{x} \vec{E}_{k}(x,y)
		\end{align*}
		Plugging this into Equation \eqref{eq:EiDeltaEj} shows that, 
		$$
			\| \vec{E}_{i}\Delta\vec{E}_{j} \|^{2}  = \sum_{k \in \llbracket 0,n \rrbracket} q_{i,j}^{k} \| \vec{E}_{k} \ket{\psi} \|^{2}
		$$
		Given $k_{0} \in \llbracket 0,n \rrbracket$, we choose $\ket{\psi}$ such that $\vec{E}_{k_{0}}\ket{\psi} \neq \vec{0}$ and $\vec{E}_{k}\ket{\psi} = \vec{0}$. It is possible since the $\vec{E}_{i}$'s are orthogonal projectors. Plugging such $\ket{\psi}$ in the above equation shows that,
		$$
		\| \vec{E}_{i}\Delta\vec{E}_{j} \|^{2} = q_{i,j}^{k_{0}} \| \vec{E}_{k_{0}}\ket{\psi} \|^{2} \geq 0
		$$
		where $i,j,k_{0} \in \llbracket 0,n \rrbracket$ are arbitrary. It concludes the proof. 
	\end{proof}	
	
	Krein parameters also appear when considering the product of $q$-numbers.

	\productKrein*

	\begin{proof}
		By Definition \ref{def:qnumber} Equation \eqref{eq:Diproj} and orthogonality of the $\vec{D}_{i}$'s whose square norm is $v_{i}$,
		$$
		q_{k}(i)q_{\ell}(i) = \frac{1}{\| \vec{D}_{i} \|^{2}} \langle  |\X| \cdot \vec{E}_{i}\circ  |\X| \cdot \vec{E}_{\ell}, \vec{D}_{i} \rangle = \frac{1}{|\X| \cdot v_{i}} \langle |\X| \cdot \vec{E}_{i}\circ |\X| \cdot \vec{E}_{\ell}, \vec{D}_{i} \rangle
		$$ 
		Now using Proposition \ref{propo:qijk},
		$$
		q_{k}(i)q_{\ell}(i) = \frac{1}{v_i}\sum_{m \in \llbracket 0,n \rrbracket} q_{k,\ell}^{m} \langle \vec{E}_{m},\vec{D}_{i} \rangle =  \frac{1}{v_i}\; \sum_{k \in \llbracket 0,n \rrbracket} q_{k,\ell}^{m} \; v_{i} \;  q_{m}(i) 
		$$
		where in the last equality we used Equation \eqref{eq:scalProdEiDi}. It concludes the proof. 
	\end{proof} \end{document}